\documentclass[twocolumn,superscriptaddress,longbibliography,prb]{revtex4-2}

\usepackage{bm,graphicx,textcomp,amssymb,amsmath,dcolumn,float,color,bbold,cancel}
\usepackage{tikz}
\usepackage{pgfplots}[compat=1.18]
\pgfplotsset{compat = newest}
\usepackage{comment}
\usetikzlibrary{plotmarks}
\newcommand{\tr}{\operatorname{tr}}

\newcommand{\ket}[1]{\left|#1\right>}
\newcommand{\bra}[1]{\left<#1\right|}

\newcommand{\nn}{\nonumber\\}

\newcommand{\bea}{\begin{eqnarray}}
\newcommand{\ea}{\end{eqnarray}}
\newcommand{\eea}{\end{eqnarray}}
\newcommand{\ord}{\,{\cal O}}

\def\h{\hat}

\newcommand{\JV}[1]{\textcolor{black}{#1}}
\newcommand{\JVn}[1]{\textcolor{black}{#1}}
\newcommand{\JN}[1]{\textcolor{black}{#1}}
\usepackage{physics,csquotes}

\usepgfplotslibrary{external} 
\usepackage{subcaption}

\begin{document}

\title{\JV{Transmission through multiple} Mott insulator--semiconductor wells}

\author{Jan Verlage} 

\affiliation{Fakult\"at f\"ur Physik and CENIDE, Universit\"at Duisburg-Essen,
  Lotharstra{\ss}e 1, 47057 Duisburg, Germany,}

%\author{Friedemann Queisser}

%\affiliation{Helmholtz-Zentrum Dresden-Rossendorf, 
%Bautzner Landstra{\ss}e 400, 01328 Dresden, Germany,}

%\affiliation{Institut f\"ur Theoretische Physik, 
%Technische Universit\"at Dresden, 01062 Dresden, Germany,}

\author{Peter Kratzer} 
\affiliation{Fakult\"at f\"ur Physik and CENIDE, Universit\"at Duisburg-Essen,
  Lotharstra{\ss}e 1, 47057 Duisburg, Germany,}

%\author{Ralf Sch\"utzhold}

%\affiliation{Helmholtz-Zentrum Dresden-Rossendorf, 
%Bautzner Landstra{\ss}e 400, 01328 Dresden, Germany,}

%\affiliation{Institut f\"ur Theoretische Physik, 
%Technische Universit\"at Dresden, 01062 Dresden, Germany,}

\date{\today}

\begin{abstract}
\JV{Weakly and strongly interacting} quantum many-body systems, namely semiconductors and Mott insulators, are combined into a layered heterostructure.
Via the hierarchy of correlations, we derive and match the propagating quasi-particle 
solutions in the different regions and calculate the transmission coefficients \JV{through these layered structures.}
%
%The resulting superlattices provide an energy filter for the injected electrons. 
%
As a proof of principle, we find the well known transmission bands of a semiconductor heterostructure.
Extending this idea to semiconductor and \JV{Mott insulator structures} we calculate the transmittance and the resonance energies. \JV{Within a phase accumulation model we find analytical expressions for the scattering phase shift. Lastly, we find transmission curves with skewness for structures with applied voltage.}

\end{abstract}

\maketitle

%%%%%%%%%%%%%%%%%%%%%%%%%%%%%%%%%%%%%%%%%%%%%%%%%%%%%%%%%%%%%%%%%%%%%%%%%%%%%%%%%%%
%%%%%%%%%%%%%%%%%%%%%%%%%%%%%%%%%%%%%%%%%%%%%%%%%%%%%%%%%%%%%%%%%%%%%%%%%%%%%%%%%%%
\section{Introduction}
%%%%%%%%%%%%%%%%%%%%%%%%%%%%%%%%%%%%%%%%%%%%%%%%%%%%%%%%%%%%%%%%%%%%%%%%%%%%%%%%%%%
%%%%%%%%%%%%%%%%%%%%%%%%%%%%%%%%%%%%%%%%%%%%%%%%%%%%%%%%%%%%%%%%%%%%%%%%%%%%%%%%%%%

Superlattices, engineered structures composed of alternating layers of different materials, have emerged as fascinating platforms for exploring novel electronic and optical properties. Their ability to exhibit unique phenomena, such as quantum confinement \cite{lu1995quantum,willatzen2004quantum,liu2011quantum} and collective behavior, has captivated researchers in various disciplines. Traditionally, superlattices have been engineered predominantly from materials with well-defined metallic or semiconducting characteristics \cite{gaylord1988semiconductor, yang2000energy, tung1996novel}, e.g., $\mathrm{GaAs/Al}_{x} \mathrm{Ga}_{1-x}\mathrm{As}$ \cite{smoliner1999biased,kan1993transmission}. Moreover, they show energy filtering over potential barriers \cite{tung1996energy, whitlow1995superlattice, summers1986variably, harness1992double}. \JV{This plays a crucial role in applications including thermoelectric devices, photovoltaics or energy storage \cite{gayner2020energy, neophytou2016modulation, mune2007enhanced}.}
In the thermoelectric setting, a skewness of the transport distribution function is desired to simultaneously improve the different constituents, enhancing the thermoelectric power \cite{zheng2022asymmetrical}.

%Energy filtering, a fundamental concept in materials science and energy harvesting, involves controlling the flow of electrons based on their energy. It plays a crucial role in various applications, including thermoelectric devices and their efficiency, photovoltaics, and energy storage devices \cite{gayner2020energy, neophytou2016modulation, mune2007enhanced}. Superlattices have already shown promise in energy filtering due to their ability to modulate band structures and exploit quantum confinement effects. In the thermoelectric setting, a skewness of the transport distribution function is wanted to simultaneously improve the different constitutents enhancing thermoelectric power \cite{zheng2022asymmetrical}.

In contrast to semiconductor characteristics, Mott insulators \cite{Avigo2020,Hubbard1963} are a class of materials characterized by their unusual behavior that defies conventional band theory. 
\JV{Standard band theory describes them as conductors, but the insulating behavior }arises from strong electron-electron interactions that result in localized, rather than itinerant, electronic states \cite{brandow1977electronic, anisimov1991band}.
Mott insulators have been the subject of intense investigation for several decades, due to their relevance to a range of phenomena, including, for example, high-temperature superconductivity \cite{lee2006doping,ju2022emergence}. \JV{In these materials, the transmission happens via quasi-particles showing distinct characteristics from usual semiconducting materials \cite{verlage2024quasi}.}

However, incorporating Mott insulators within the superlattice architecture,
e.g. as ultra-thin silicide layers in silicon \cite{Qian-PRL-2006,Lin-NanoLett-20}, introduces an additional dimension of control through strong electron correlations. \JV{The strength of the Coulomb interaction between the electrons changes the size of the Mott gap. This effectively influences the transmission of electrons passing through these layers as doublon-holon quasi-particles. This allows for a better tailoring of the transmission characteristics compared to structures that lack this degree of freedom, and the Mott insulating layers introduce an additional skewness of the transmission favorable in thermoelectric applications.}

\JN{One material class suitable for this heterostructure design are perovskites like TiF$_3$~\cite{sheets2023mott} and LaTiO$_3$~\cite{maznichenko2024fragile}, especially the interfaces $\mathrm{LaAlO}_3/\mathrm{SrTiO}_3$~\cite{ohtomo2004high,thiel2006tunable}, $\mathrm{LaAlO}_3/\mathrm{SrTiO}_3$~\cite{pentcheva2006charge}, $\mathrm{LaTiO}_3/\mathrm{KTaO}_3$~\cite{maznichenko2024fragile,Maznichenko2024} or $\mathrm{LaTiO}_3/\mathrm{SrTiO}_3$~\cite{santana2019electron,ishida2008origin,okamoto2004electronic}. The correlation strength in these perovskites can be tuned by varying the central atom, and their interfaces  can be grown with high precision.}
\JN{Alternatively, stacked heterostructures bound by van der Waals forces \cite{castellanos2022van} facilitate the integration of  various components such as single-band Mott insulators~\cite{PhysRevX.13.041049}, the Mott insulating state of FeSe~\cite{kang2024mott}, or layered Mott insulating materials like FeP$X_3$ ($X =$ S, Se)~\cite{jin2022mott}, alongside semiconducting van der Waals materials.}

\JV{Usually the Mott behavior is treated within a DMFT framework which takes the individual layers as infinite dimensional \cite{Okamoto2004,okamoto2007nonequilibrium}. In this work, we use an approach via the } hierarchy of correlations \cite{Queisser2014,navez2010emergence,queisser2019boltzmann}.
This method provides a powerful tool for describing both the weakly interacting semiconductors and strongly interacting Mott insulators on the same footing while maintaining spatial resolution.

The article is organized as follows; we first introduce the Hubbard model used as a theoretical description for the different material types \JV{as well as} the hierarchy of correlations, the proof of concept involving superlattices constructed by different semiconductors is calculated afterwards and lastly the extension to semiconductor-Mott insulator heterostructures is made.

%%%%%%%%%%%%%%%%%%%%%%%%%%%%%%%%%%%%%%%%%%%%%%%%%%%%%%%%%%%%%%%%%%%%%%%%%%%%%%%%%%%
%%%%%%%%%%%%%%%%%%%%%%%%%%%%%%%%%%%%%%%%%%%%%%%%%%%%%%%%%%%%%%%%%%%%%%%%%%%%%%%%%%%
\section{Fermi-Hubbard Model}
%%%%%%%%%%%%%%%%%%%%%%%%%%%%%%%%%%%%%%%%%%%%%%%%%%%%%%%%%%%%%%%%%%%%%%%%%%%%%%%%%%%
%%%%%%%%%%%%%%%%%%%%%%%%%%%%%%%%%%%%%%%%%%%%%%%%%%%%%%%%%%%%%%%%%%%%%%%%%%%%%%%%%%%
In order to describe both the strongly correlated Mott insulating and the semiconductor sites we employ a tight-binding model, namely the Fermi Hubbard model~\cite{Hubbard1963,arovas2022hubbard,qin2022hubbard}
\bea
\label{Fermi-Hubbard}
\hat H=-\frac1Z\sum_{\mu\nu s} T_{\mu\nu} \hat c_{\mu s}^\dagger \hat c_{\nu s} 
+\sum_\mu U_\mu \hat n_\mu^\uparrow\hat n_\mu^\downarrow
+\sum_{\mu s}V_\mu\hat n_{\mu s}  
\,.\;
\ea
This Hamiltonian describes the relevant band of the semiconductor, either valence or conduction band, with $U_\mu \equiv 0$, and the Mott insulator with non-zero on-site repulsion $U_\mu \neq 0$. \JV{Even if there is a small interaction in the semiconductor this is absorbed into an effective on-site potential $V_\mu$ similar to Fermi liquid theory \cite{landau1957theory,solovyev2017renormalized}.}
As usual, $\hat c_{\mu s}^\dagger$ and $\hat c_{\nu s}$ denote the fermionic 
creation and annihilation operators at the lattice sites $\mu$ and $\nu$ 
with spin $s\in\{\uparrow,\downarrow\}$ and $\hat n_{\mu s}$ are the 
associated number operators. 
The hopping matrix $T_{\mu \nu}$ encodes both the adjacency and the hopping strength. In the following, it is assumed in the form of nearest neighbor hopping and zero otherwise. Moreover, we assume a uniform hopping strength throughout the entire heterostructure, but the results can be generalized to non-uniform hopping in a straightforward manner.
The coordination number $Z$ counts the nearest neighbors $\nu$ for any site $\mu$ and is assumed large, $Z \gg 1$.
\JV{The different regions are differentiated by the two parameters, $U_\mu$ and $V_\mu$. The Mott insulator is characterized by a large on-site repulsion $U \gg T$, creating the upper and lower Hubbard band, and zero on-site potential $V_\mu \equiv 0$. The semiconductor has an on-site potential $V_\mu \neq 0$, because only the relative positions of the semiconducting and Hubbard bands matter, and no on-site repulsion.}
\subsection{Hierarchy of Correlations}
In the following \JV{we are interested in the propagation of electrons through heterostructures. Within the Mott insulating layers this propagation happens by quasi-particles, namely doublons $\ket{\uparrow \downarrow}$ in the upper Hubbard band and holons $|0\rangle$ in the lower Hubbard band on top of a half-filled background \cite{verlage2024quasi}. This is similar to the Hubbard-I approximation \cite{Hubbard1963,hubbard1965electron}.} To this end, we consider the reduced density matrices for one $\h \rho_\mu$, two $\h \rho_{\mu \nu}$ and more lattice sites. This is done by tracing out all other lattice sites but one, $\h \rho_\mu = \tr_{\cancel{\mu}}\h \rho$. Additionally, we split up the correlations, $\h \rho_{\mu \nu}=\h \rho_\mu \h \rho_\nu + \h \rho_{\mu \nu}^\mathrm{corr}$ and so on. From this starting point, we employ an expansion into powers of the inverse coordination number $1/Z$ based on the $Z \gg 1$ assumption. Doing so, higher order correlations are successively suppressed. While the two-site correlations scale as $\h \rho_{\mu \nu}^\mathrm{corr}=\mathcal{O}(1/Z)$, the three-sites ones scale as $\h \rho_{\mu \nu \lambda}^\mathrm{corr}=\mathcal{O}(1/Z^2)$ and so on. This hierarchy yields an iterative scheme for the correlations.
\JV{Mathematically speaking, we start} from the exact evolution equations for these correlations 
($\hbar=1$)
\bea
\label{evolution}
i\partial_t \hat\rho_\mu 
&=& 
F_1(\hat\rho_\mu,\hat\rho_{\mu\nu}^{\rm corr})
\,,\nn
i\partial_t \hat\rho_{\mu\nu}^{\rm corr} 
&=& 
F_2(\hat\rho_\mu,\hat\rho_{\mu\nu}^{\rm corr},\hat\rho_{\mu\nu\lambda}^{\rm corr})
\,,
%\nn
%i\partial_t \hat\rho_{\mu\nu\lambda}^{\rm corr} 
%&=& 
%F_3(\hat\rho_\mu,\hat\rho_{\mu\nu}^{\rm corr},\hat\rho_{\mu\nu\lambda}^{\rm corr},
%\hat\rho_{\mu\nu\lambda\kappa}^{\rm corr})
%\,,
\ea
\JV{that we get from the Heisenberg equation.}
We now approximate them using the expansion into the inverse powers of the coordination number. 
The functions $F_n$ are determined by the Hamiltonian of the system Eq.~\ref{Fermi-Hubbard}, or more precisely by the commutator with the reduced density matrix.
Starting the iterative scheme to lowest order $\mathcal{O}(Z^0)$, the evolution equation for the on-site density matrix is given by
$i\partial_t \hat\rho_\mu = F_1(\hat\rho_\mu,0)+\ord(1/Z)$.  
Its zeroth-order solution $\hat\rho_\mu^0$ yields the mean-field background as the starting 
point for the higher orders. 
For the sake of the Mott insulator description at half-filling and in the regime of large repulsion $U \gg T$, the mean-field \textit{ansatz} we choose is
\bea
\label{mean-field}
\hat\rho_\mu^0
=
\frac{\ket{\uparrow}_\mu\!\bra{\uparrow}+\ket{\downarrow}_\mu\!\bra{\downarrow}}{2}
\,.
\ea
%This mean-field state is accompanied by virtual hopping processes that create doublons $\ket{\uparrow \downarrow}$ and holons $|0\rangle$ of probability $\mathcal{O}(J^2/U^2)$ that could be included. Due to the strong suppression of this in the parameter regions considered, will not do this here.
There is one particle per site without any spin ordering. 
The semiconductor is represented by 
$\hat\rho_\mu^0=\ket{\uparrow\downarrow}_\mu\!\bra{\uparrow\downarrow}$
for the valence band and by 
$\hat\rho_\mu^0=\ket{0}_\mu\!\bra{0}$ for the conduction band
(at zero temperature). 
Note that the two cases are related to each other via particle-hole duality. For more details see \cite{Queisser2014,verlage2024quasi}.

\subsection{Quasi-Particle Operators}
To first order $\mathcal{O}(Z^{-1})$, \JV{the correlation functions obey} 
$i\partial_t \hat\rho_{\mu \nu}^\mathrm{corr} = F_2(\hat\rho_\mu^0,\hat\rho_{\mu \nu}^\mathrm{corr})+\ord(1/Z^2)$
on top of the mean-field background.
For a more detailed view see, e.g, \cite{Queisser2014,navez2010emergence,queisser2019boltzmann}.

To analyze the evolution of the quasi-particles
it is beneficial to introduce the quasi-particle and hole operators 
\bea
\hat c_{\mu s I}=\hat c_{\mu s}\hat n_{\mu\bar s}^I=
\left\{
\begin{array}{ccc}
 \hat c_{\mu s}(1-\hat n_{\mu\bar s}) & {\rm for} & I=0 
 \\ 
 \hat c_{\mu s}\hat n_{\mu\bar s} & {\rm for} & I=1
\end{array}
\right.
\,,
\ea
where $\bar s$ denotes the spin state opposite to $s$. They create doublons ($I=1$) or holons ($I=0$).
Note that these operators are approximately, but not exactly equal to the 
quasi-particle creation and annihilation operators for holons and doublons, 
see, e.g. \cite{Avigo2020}. \JV{The idea of such effective operators that better describe the physical excitations goes back to Hubbard X operators \cite{hubbard1965electron,ovchinnikov2004hubbard} and composite operators \cite{mancini2004hubbard}.}

In terms of these operators, 
we are able to write down the evolution equation for the correlation functions $\langle\hat c^\dagger_{\mu s I}\hat c_{\nu s J}\rangle^{\rm corr}$. They consist of a homogeneous part coupling the different correlations and stationary source terms, that are not needed to describe the propagation of the quasi-particles . The source terms can be used to calculate ground state correlations. Moreover, it can be shown that the factorization
\bea
\langle\hat c^\dagger_{\mu s I}\hat c_{\nu s J}\rangle^{\rm corr}
=p_{\mu s I}^* p_{\nu s J}
\,,
\ea
where the factors $p_{\nu s I}$ obey the simple equation 
\bea
\label{eq:6}
\left(i\partial_t-U^I_\mu-V_\mu\right)p_{\mu s I}=
\frac{-1}{Z}\sum_{\nu J} T_{\mu\nu} \langle\hat n_{\mu\bar s}^I\rangle^0 p_{\nu s J},
\ea
yields the same dynamics, for details see \cite{verlage2024quasi}.
 Furthermore, we assume a highly symmetric lattice such as a hyper-cubic one, to apply a Fourier transformation of the dependency parallel to the interfaces in the system. It reads
\begin{equation}
	\begin{aligned}
		p_{\mu s I}&=\frac{1}{\sqrt{N^\parallel}}\sum_{\mathbf{k}^\parallel}p_{n,\mathbf{k}^\parallel,s}^Ie^{i  \mathbf{k}^\parallel \cdot\mathbf{x}_\mu^\parallel}\\
		T_{\mu\nu}&=\frac{Z}{N^\parallel}\sum_{\mathbf{k}^\parallel} T_{m,n,\mathbf{k}^\parallel}e^{i \mathbf{k}^\parallel \cdot\left(\mathbf{x}_\mu^\parallel-\mathbf{x}_\nu^\parallel\right)} \,.
	\end{aligned}
\end{equation}
and for isotopic nearest neighbor hopping  $T^\|_n=T^\perp_{n,n-1}=T$ the components read
\begin{equation}
\begin{aligned}
	T_{m,n,\mathbf{k}^\parallel}&=\frac{T^\parallel_{\mathbf{k}^\parallel}}{Z}\delta_{m,n}+
	\frac{T}{Z}
	(\delta_{n,n-1}+\delta_{n,n+1})\\
	T_{\mathbf{k}^\parallel}^\parallel&=2 T \sum_{x_i}\cos(p_{x_i}^\parallel) \equiv Z T_\mathbf{k}^{\|}\,,
\end{aligned}	
\end{equation}
with the hopping contribution $T_{\bf k}^\|$. \JN{In this hyper-cubic lattice, there are $Z=4$ nearest neighbors in a two-dimensional lattice and $Z=6$ in a three-dimensional one.}
 From Eq.~\ref{eq:6} we read of that the different spin sectors decouple, therefore we drop the $s$ and ${\bf k}^\|$ indices 
for readability and find
\bea
\label{difference}
\left(E-U_n^I-V_n\right)p_n^I
+{\color{black}\langle\hat n_{n}^I\rangle^0
\sum_J T_{\bf k}^\| p_n^J}
=
\nn
-T\frac{\langle\hat n_{n}^I\rangle^0}{Z}
\sum_J \left(p_{n-1}^J+p_{n+1}^J\right) 
\,,
\ea
where $n\in\mathbb Z$ now just labels the lattice sites perpendicular 
to the interface. 
These equations show a linear dependency of \JV{quasi-}particle $p_n^1$ and hole $p_n^0$ solutions in the Mott insulator while in the semiconductor one of them is identically zero, in the 
valence or conduction band, respectively. 
 \JN{The assumption of a hyper-cubic lattice is not only mathematically simplifying the analysis, but also guided by particularly perovskite like Mott insulators. These show Mott insulating behavior on three-dimensional cubic lattices with $Z=6$ nearest neighbors, e.g., TiF$_3$  \cite{sheets2023mott} and LaTiO$_3$ \cite{maznichenko2024fragile}.} 
 \JN{However, we stress that the hierarchy of correlations works for any lattice structure, such that Mott insulators on (quasi-)two-dimensional triangular lattices with $Z=6$ neighbors \cite{PhysRevB.94.161105, tomeno2020triangular, PhysRevLett.99.256403} could also be described.}

In those regions of constant $U_n$ or $V_n$ we solve Eq.~\ref{difference}
by \bea
\label{ansatz} 
p_n^I=\alpha^I e^{i\kappa n}+\beta^I e^{-i\kappa n}
\,.
\ea
This yields an effective wave number $\kappa$ perpendicular to the interface.
The propagation within the semiconductor with a vanishing repulsion 
$U_n=0$ and a constant on-site potential $V_n=V$, is governed by the 
effective wave number
\bea
\label{semiconductor}
\cos\kappa_\mathrm{semi}=\frac{Z}{2T}\left[V-E-T_{\bf k}^\|\right]
\,.
\ea
This is a band-like dispersion relation with $E=V-T_\mathbf{k}$.
The Mott insulator, however, corresponds to a large constant on-site repulsion $U_n=U\gg T$ with a vanishing potential $V_n=0$. 
In this case, the effective wave number reads 
\bea
\label{Mott}
\cos\kappa_\mathrm{Mott}=\frac{Z}{2T}\left[\frac{E(U-E)}{E-U/2}-T_{\bf k}^\|\right]
\,.
\label{kappaMott}
\ea
\JV{The corresponding dispersion in the Mott insulator reads $E=\frac{1}{2}\left(U-T_{\bf k} \pm \sqrt{U^2+T_{\bf k}^2} \right)$ with the kinetic energy contribution $T_{\bf k}=T_{\bf k}^\| + \frac{2T}{Z}\cos\kappa_\mathrm{Mott}$.}
\JV{This dispersion relation falls in line with the ones obtained by other methods like  Hubbard-I \cite{hubbard1965electron}, Roth's two pole approximation \cite{roth1969electron} or composite operator methods \cite{avella2014hubbard}}.
\JN{In DMFT the $Z\to \infty$ assumption fixes the effective bandwidth of $T/\sqrt{Z}$~\cite{RevModPhys.68.13,PhysRevB.45.6479}. Within the hierarchy of correlations, signatures of this effective bandwidth can be seen even in finite dimension. 
%PK Ist der folgende Satz offensichtlich, 
%PK oder ist das eine spezielle Beobachtung (von wem)?
For a fixed effective bandwidth, the asymptotic magnitude of double occupations becomes independent of the dimension after a quantum quench~\cite{PhysRevB.109.195140}. This suggest the stability of the hierarchy of correlations. Moreover, the %influence of 
%PK Vorschlag
the role of dimensionality for 
higher-order corrections (and hence scaling with the power of $Z$) is investigated in Ref.~\onlinecite{PhysRevB.109.195140} for cubic lattices.
%in several dimensions. 
Furthermore, it would be possible to incorporate second-order effects via renormalization of the hopping parameter~\cite{Queisser2014} if necessary, thus ensuring that out predictions remain valid across these systems.
%PK vielleicht rauslassen
%Right now, these higher order calculations only exist for homogeneous systems and are subject of future works.
}
For more details about this and a comparison between the hierarchy of correlations and DMFT see \cite{verlage2024quasi}.
In the strongly interacting regime $U\gg T$ considered here, 
real solutions for $\kappa$ exist only for small $E \approx 0$ (lower Hubbard band) 
or for $E\approx U$ (upper Hubbard band).

The dispersion relations found above for the doublons and holons are the starting point to calculate their transmission through different types of heterostrucures, made up either by different semiconductors or semiconductors and Mott insulators.

%%%%%%%%%%%%%%%%%%%%%%%%%%%%%%%%%%%%%%%%%%%%%%%%%%%%%%%%%%%%%%%%%%%%%%%%%%%%%%%%%%%
%%%%%%%%%%%%%%%%%%%%%%%%%%%%%%%%%%%%%%%%%%%%%%%%%%%%%%%%%%%%%%%%%%%%%%%%%%%%%%%%%%%
\section{Semiconductor Heterostructure}
\label{section:semis}
%%%%%%%%%%%%%%%%%%%%%%%%%%%%%%%%%%%%%%%%%%%%%%%%%%%%%%%%%%%%%%%%%%%%%%%%%%%%%%%%%%%
%%%%%%%%%%%%%%%%%%%%%%%%%%%%%%%%%%%%%%%%%%%%%%%%%%%%%%%%%%%%%%%%%%%%%%%%%%%%%%%%%%%
\begin{figure}
    \centering
   	\begin{tikzpicture}[scale=0.7, transform shape]
			% Parameters
		\def\squareSize{1cm} % Size of each square
		\def\spacing{0.2cm} % Space between squares
		\def\rows{2} % Number of rows
		\def\cols{5} % Number of columns
		\def\colsm{4}
		% Loop to draw the grid
		
		\foreach \i in {0,...,\rows} {
			\foreach \j in {0,...,2} {
				% Draw the square
				\draw[rounded corners=5pt, red] 
				(\j * \squareSize + \j * \spacing, -\i * \squareSize - \i * \spacing) 
				rectangle ++(\squareSize, \squareSize);}}		
		
		\foreach \i in {0,...,\rows} {
			\foreach \j in {3,...,5} {
				% Draw the square
				\draw[rounded corners=5pt, blue] 
				(\j * \squareSize + \j * \spacing, -\i * \squareSize - \i * \spacing) 
				rectangle ++(\squareSize, \squareSize);}}	
		
		\foreach \i in {0,...,\rows} {
			\foreach \j in {6,...,8} {
				% Draw the square
				\draw[rounded corners=5pt, green] 
				(\j * \squareSize + \j * \spacing, -\i * \squareSize - \i * \spacing) 
				rectangle ++(\squareSize, \squareSize);}}

		\def\cornerRounding{5pt} % Rounding of corners
		
		%% at the top
		\foreach \i in {0,...,2} {
			\draw[rounded corners=\cornerRounding, red]  (\i * \squareSize+\i * \spacing,1.5*\squareSize+\spacing) -- ++(0, -0.5*\squareSize) -- ++(\squareSize, 0) -- ++(0, 0.5*\squareSize) ;
			\pgfmathsetmacro{\imone}{\i - 1}
			\pgfmathsetmacro{\ipl}{\i + 1}
			\draw[rounded corners=\cornerRounding,red] 
			(\ipl*\squareSize+\i*\spacing,-3*\squareSize-0.5*\spacing) -- ++(0, 0.5*\squareSize) -- ++(-\squareSize, 0) -- ++(0, -0.5*\squareSize) ;
		}
		
		%% at the top
		\foreach \i in {3,...,5} {
			\draw[rounded corners=\cornerRounding, blue]  (\i * \squareSize+\i * \spacing,1.5*\squareSize+\spacing) -- ++(0, -0.5*\squareSize) -- ++(\squareSize, 0) -- ++(0, 0.5*\squareSize) ;
			\pgfmathsetmacro{\imone}{\i - 1}
			\pgfmathsetmacro{\ipl}{\i + 1}
			\draw[rounded corners=\cornerRounding,blue] 
			(\ipl*\squareSize+\i*\spacing,-3*\squareSize-0.5*\spacing) -- ++(0, 0.5*\squareSize) -- ++(-\squareSize, 0) -- ++(0, -0.5*\squareSize) ;
		}
		
		%% at the top
		\foreach \i in {6,...,8} {
			\draw[rounded corners=\cornerRounding, green]  (\i * \squareSize+\i * \spacing,1.5*\squareSize+\spacing) -- ++(0, -0.5*\squareSize) -- ++(\squareSize, 0) -- ++(0, 0.5*\squareSize) ;
			\pgfmathsetmacro{\imone}{\i - 1}
			\pgfmathsetmacro{\ipl}{\i + 1}
			\draw[rounded corners=\cornerRounding,green] 
			(\ipl*\squareSize+\i*\spacing,-3*\squareSize-0.5*\spacing) -- ++(0, 0.5*\squareSize) -- ++(-\squareSize, 0) -- ++(0, -0.5*\squareSize) ;
		}

		%% at the right
		\draw[rounded corners=\cornerRounding, blue] 
		(9.5*\squareSize+9*\spacing,-2.5*\squareSize-.5*\spacing) -- ++(-0.5*\squareSize,0) -- ++(0,-0.5*\squareSize); 		
		\draw[rounded corners=\cornerRounding, blue] 
		(9.5*\squareSize+9*\spacing,1*\squareSize+1*\spacing) -- ++(-0.5*\squareSize,0) -- ++(0,0.5*\squareSize);

		\draw[rounded corners=\cornerRounding, blue] 
		(9.5*\squareSize+9*\spacing,-1.5*\squareSize+1.5*\spacing) -- ++(-0.5*\squareSize,0) -- ++(0,\squareSize) -- ++(0.5*\squareSize,0) ;
		\draw[rounded corners=\cornerRounding, blue] 
		(9.5*\squareSize+9*\spacing,-2.5*\squareSize+.5*\spacing) -- ++(-0.5*\squareSize,0) -- ++(0,\squareSize) -- ++(0.5*\squareSize,0) ;
		\draw[rounded corners=\cornerRounding, blue] 
		(9.5*\squareSize+9*\spacing,-0.5*\squareSize+2.5*\spacing) -- ++(-0.5*\squareSize,0) -- ++(0,\squareSize) -- ++(0.5*\squareSize,0) ;
		
		%% at the left
		\draw[rounded corners=\cornerRounding, blue] 
		(-1*\squareSize+1.5*\spacing,-2.5*\squareSize-0.5*\spacing) -- ++(0.5*\squareSize,0) -- ++(0,-0.5*\squareSize) ;	
		\draw[rounded corners=\cornerRounding, blue] 
		(-1*\squareSize+1.5*\spacing,1*\squareSize+1.*\spacing) -- ++(0.5*\squareSize,0) -- ++(0,0.5*\squareSize) ;		
		
		\draw[rounded corners=\cornerRounding, blue] 
		(-1*\squareSize+1.5*\spacing,-1.5*\squareSize+0.5*\spacing) -- ++(0.5*\squareSize,0) -- ++(0,-\squareSize) -- ++(-0.5*\squareSize,0) ;
		
		\draw[rounded corners=\cornerRounding, blue] 
		(-1*\squareSize+1.5*\spacing,-0.5*\squareSize+1.5*\spacing) -- ++(0.5*\squareSize,0) -- ++(0,-\squareSize) -- ++(-0.5*\squareSize,0) ;
		
		\draw[rounded corners=\cornerRounding, blue] 
		(-1*\squareSize+1.5*\spacing,.5*\squareSize+2.5*\spacing) -- ++(0.5*\squareSize,0) -- ++(0,-\squareSize) -- ++(-0.5*\squareSize,0) ;
		%% vertical lines
		\foreach \j in {-1,...,9} {
			\draw[gray!40,line width=1pt] (0.5*\squareSize+\j * \squareSize + \j * \spacing,0) -- ++(0,-\spacing);
			\draw[gray!40,line width=1pt] (0.5*\squareSize+\j * \squareSize + \j * \spacing,-\squareSize - \spacing) -- ++(0,-\spacing);
		}
		\foreach \j in {-1,...,9} {
			\draw[gray!40,line width=1pt] (0.5*\squareSize+\j * \squareSize + \j * \spacing,1*\squareSize +1*\spacing) -- ++(0,-\spacing);
			\draw[gray!40,line width=1pt] (0.5*\squareSize+\j * \squareSize + \j * \spacing,-2*\squareSize - 2*\spacing) -- ++(0,-\spacing);
		}
		%% horizontal lines
		\foreach \j in {-2,...,0} {
			\foreach \i in {0,...,9}{
				\draw[gray!40,line width=1pt] (\i*\squareSize+\i*\spacing,0.5*\squareSize+\j * \squareSize + \j * \spacing) -- ++(-\spacing,0);
			}
		}
		
		\foreach \j in {0,...,9} {
			\draw[gray!40,line width=1pt] (\j * \squareSize+\j * \spacing, -3.*\squareSize + 0.5*\spacing) -- ++(-\spacing,0);	
			\draw[gray!40,line width=1pt] (\j * \squareSize+\j * \spacing, 1.5*\squareSize + 0.5*\spacing) -- ++(-\spacing,0);	
		}

		\draw[fill=white, draw=white] (1.43*\squareSize+0*\spacing,-3.3) rectangle ++(0.25*\squareSize+1.5*\spacing,5);
		\draw[fill=white, draw=white] (4.43*\squareSize+3*\spacing,-3.3) rectangle ++(0.25*\squareSize+1.5*\spacing,5);
		\draw[fill=white, draw=white] (7.43*\squareSize+6*\spacing,-3.3) rectangle ++(0.25*\squareSize+1.5*\spacing,5);
		
		\node[blue] at (0*\squareSize-3*\spacing,0.5) {\huge$V$};
		\node[blue] at (4*\squareSize+2*\spacing-1*\spacing,0.5) {\huge$V$};
		\node[blue] at (10*\squareSize+6*\spacing-0.2*\spacing,0.5) {\huge$V$};
		\node[red] at (1*\squareSize-2*\spacing,0.472) {\huge$V_1$};
		\node[green] at (7*\squareSize+3*\spacing,0.475) {\huge$V_2$};
		
		\node[black] at (0*\squareSize-2*\spacing,-3.5) {\large$n=0$};
		\node[black] at (3*\squareSize-0.5*\spacing,-3.5) {\large$n=a$};
		\node[black] at (6*\squareSize+2.5*\spacing,-3.5) {\large$n=b$};
		\node[black] at (9*\squareSize+5.5*\spacing,-3.5) {\large$n=c$};
		
	\end{tikzpicture}
    \caption{(Color online) \JN{Two-dimensional} schematic of the system \JN{signifying the whole system in 2D, or  a slice from a three-dimensional lattice}; different semiconductors with potential $V$, $V_1$ and $V_2$ are \JV{combined} to a heterostructure.}
    \label{fig:system}
\end{figure}
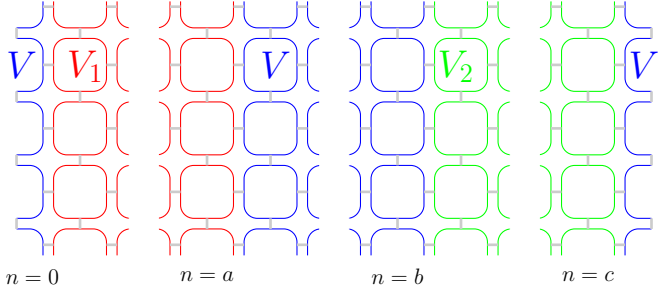
\JV{As proof of principle, we reproduce some well known results \cite{yang2000energy} using the hierarchy of correlations. We consider a pure semiconductor structure matching previous studies employing the transfer matrix method \cite{yang2000energy}, which often involve superlattices with numerous wells.} The schematic representation of our structure is presented in Fig.~\ref{fig:system}. \JV{There is one type of semiconducting material with on-site potential $V$ functioning as the leads for sites $n \in (-\infty,0]$ and $n \in [c+1,\infty)$, as well as the connecting layer $n \in [a+1,b]$. It connects two different layers with $V_1$ in $n \in [1,a]$ and $V_2$ in $n \in [b+1,c]$.}
As mentioned, we assume an uniform hopping strength $T$ across the entire lattice, although this restriction could be lifted straightforwardly. In the following analysis, the solution for the semiconductor given by Eq.~\ref{difference} is denoted as $s_\pm^n = e^{\pm i \kappa_\mathrm{semi} n}$, with the wavevector defined in Eq.~\ref{semiconductor}. Additionally, $s_{j\pm}$, with $j=1,2$, provide the solution in the intermediate regions characterized by potentials $V_1$ and $V_2$.
Without normalization, the \textit{ansatz} we use is 
\begin{equation}
\label{eq:ansatzsemistructure}
    p_n^0 = \begin{cases}
    s_+^n+R s_-^n & n \leq 0, \\
   A s_{1+}^n+B s_{1-}^n & \ 0<n \leq a, \\
    C s_+^n + D s_-^n & a < n \leq b, \\
    E s_{2+}^n+ F s_{2-}^n,  & b < n \leq c, \\
    \mathcal{T} s_+^n & c < n,
    \end{cases}
\end{equation}
as $s_+$ describes right propagating and $s_-$ left propagating solutions.
\JV{From Fig.~\ref{fig:system} and Eq.~\ref{difference} we read of boundary equations at the four interfaces and solve for the transmission amplitude $\mathcal{T}$
(see App.~\ref{app:semis} for the boundary conditions and the expression for the transmission amplitude).} \\

%The transmission amplitude does not just depend on the energy but also on the parallel momentum $\mathcal{T}=\mathcal{T}(E,k^\|)$. Due to this, there is not just a filtering of the energy but also of parallel momentum for a given incident energy $E$ (that will not be discussed here) \cite{lu2017controllable,liu2017tunable,zhang2019wave,tang2022electron}. 
%The efficiency of this wave-vector filtering is quantified by the derivative $\mathrm{wfe}=\frac{\partial \mathcal{T}(E,k^\|)}{\partial k^\|}$ \cite{lu2017controllable,liu2017tunable,zhang2019wave,tang2022electron}.

\begin{figure}[t]
    \centering
   	\begin{tikzpicture}
		\begin{axis}[
			xmin = .7, xmax = 1.,
			ymin = 0, ymax = 1.3,
			xtick distance = .1,
			%ytick distance = 0.1,
			%grid = both,
			minor tick num = 1,
			major grid style = {lightgray},
			minor grid style = {lightgray!25},
			width = \linewidth,
			height = 0.7\linewidth,
			xlabel = {$E/V$},
			ylabel = {$\frac{j^\mathrm{trans}}{j^\mathrm{in}}$},
			legend cell align = {right},
            legend style={at={(0.2,.95)},anchor=west},
            %legend columns=3
           % ymode = log
			]

			\addplot[
			smooth,
			thin,
            dotted,
          %  blue
			] table[x index=0, y index=1] {pics/semis_all.dat};

            \addplot[
			smooth,
			thin,
            black
			] table[x index=0, y index=2] {pics/semis_all.dat};

             \addplot[
			smooth,
			thin,
            dashed
			] table[x index=0, y index=3] {pics/semis_all.dat};

			\legend{$V_1/V=0.9$ $V_2/V=0.9$, $V_1/V=0.9$ $V_2/V=1.1$,$V_1/V=1.1$  $V_2/V=1.2$}
					
		\end{axis}

	\end{tikzpicture}
    \caption{Transmission \JV{as a function of electron energy} through the semiconductor heterostructure \JV{of various band-offsets.} We use $Z=4$ for a two-dimensional problem and $T/V=0.3$, $a=7$,$b=10$, $c=17$. The band width of the \enquote{leads} is given by the plot range.}
    \label{fig:trans_semis}
\end{figure}
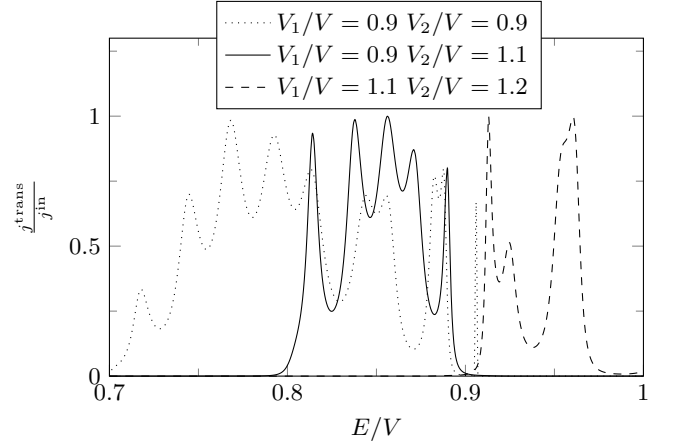

The effective wave number Eq.~\ref{semiconductor} for the semiconducting layers shows that evanescent solutions have $\abs{\kappa_s} \propto \ln{V}$. The stronger the band mismatch, the stronger the suppression of tunneling.
In combination with the injected energy of the plane wave $E \in [V,V-T]$ 
and 
a sufficiently large filtering layer $V_1$ and $V_2$, all energies $E>\mathrm{min}\{V_1,V_2\}$ get filtered out if $V_1,V_2<V$. In Fig.~\ref{fig:trans_semis}, this is shown exemplarily for $V_1=V_2=0.9V$ (dotted curve) and a fixed $k^\|$; for $E<V_1$ there is {\JVn{nearly }}zero transmission. {\JVn{There are exponentially small tunnel currents.}}
Otherwise, for $V_1 < V < V_2$ there might be an overlap in the bands around $V_1$. Consequently, everything outside of this overlap gets filtered out, see solid curve in Fig.~\ref{fig:trans_semis}. 
Using $V_1,V_2 > V$ on the other hand, everything below $E<\mathrm{max}\{V_1,V_2\}-T$ gets filtered out, see the dashed curve in the same figure.
In all three difference scenarios, there are a lot of oscillations in the transmission. \JN{Changing the layer thickness by changing $a$, $b$ and $c$ does not qualitatively change the results.}
{\JVn{By including more and more structure to create a superlattice \cite{yang2000energy} these oscillations can be smoothed out and we would find real transmission minibands, because there is more constructive interference in these superlattices than in the small system here.}}
%This behavior can be improved by including more and more structure to create a superlattice and \cite{yang2000energy} benefit from constructive interference. 

{\JVn{If the hopping is not isotropic, it is possible to achieve the same energy filtering with just two types of semiconductors. The filtering regions ought to have smaller hopping than the leads. This then results in symmetric transmission minibands. With isotropic hopping this system is only capable of filtering out energies above a certain value.}}

\subsection{Smaller structure -- non-isotropic hopping}
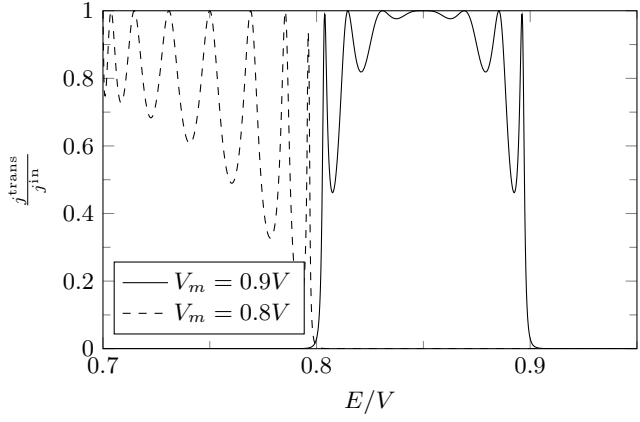
\begin{figure}[t]
    \centering
   	\begin{tikzpicture}
		\begin{axis}[
			xmin = .7, xmax = .95,
			ymin = 0, ymax = 1.,
			xtick distance = .1,
			%ytick distance = 0.1,
			%grid = both,
			minor tick num = 1,
			major grid style = {lightgray},
			minor grid style = {lightgray!25},
			width = \linewidth,
			height = 0.7\linewidth,
			xlabel = {$E/V$},
			ylabel = {$\frac{j^\mathrm{trans}}{j^\mathrm{in}}$},
			legend cell align = {right},
            legend style={at={(0.02,.15)},anchor=west},
            %legend columns=3
           % ymode = log
			]

			\addplot[
			smooth,
			thin,
			] table[x index=0, y index=1] {nonisohopping_v1_j03_a7_vm9_jm01_kp0_z4.dat};

            		\addplot[
			smooth,
			thin,
            dashed
			] table[x index=0, y index=1] {nonisohopping_v1_j03_a7_vm8_jm01_kp0_z4.dat};

       \legend{$V_m=0.9V$,$V_m=0.8V$}
			
		\end{axis}

	\end{tikzpicture}
    \caption{Transmission as a function of energy through the structure with two types of semiconductors. We use $Z=4$ for a two-dimensional problem and $T/V=0.3$ and $a=7$. The gate potential is $V_m/V=0.9$ (solid) and $V_m/V=0.8$ (dashed) and the hopping within the gate is $T_m/V=0.1$.}
    \label{fig:trans_semis_smallersystem}
\end{figure}

\JVn{As mentioned, in the case of non-isotropic hopping the energy filtering effect and minibands can be achieved by a structure consisting of only two different semiconductors; one acting as the leads and another one with smaller hopping as the gate. With a similar \textit{ansatz} as for the bigger structure with $s_\pm$ and $m_\pm$ describing the solutions in the leads and the gate, respectively, we find the transmission amplitude
\begin{equation}
    \mathcal{T}=\frac{\left(m^2-1\right) \left(s^2-1\right) m^{a+1} s^{-a-1}}{(m s-1)^2-m^{2 a+2} (m-s)^2}.
\end{equation}
It is exemplarily shown in Fig.~\ref{fig:trans_semis_smallersystem} for sample parameters $V_m/V=0.9$ and $T/T_m=3$. By shifting the gate potential and thus the band overlap this setup can filter all energies above or below a certain energy or produce transmission minibands. These can only be symmetric.
}

\section{Mott--Semiconductor structure}
\JV{Having successfully demonstrated our method for a well studied case of a semiconductor structure,} our focus now shifts toward investigating a Mott insulator-semiconductor heterostructure. In this new configuration, semiconducting leads, serving as source and drain terminals, encase a Mott insulating material characterized by an on-site repulsion parameter denoted as $U$. Within this insulating region, there exists another semiconductor segment controlled by a gate voltage $V_m$. A schematic representation of this structure is provided in Fig.~\ref{fig:onegate}. \JV{Fig.~\ref{fig:band_structure_one} shows the band edges of the lead, drain and gate relative to the Hubbard bands of the Mott insulating layers}. Notably, the application of a gate voltage results in a shift of the (red) band within the middle region, denoted as $V_m-T_\mathbf{k}$, thereby modifying the transmission characteristics.
\begin{figure*}
\centering
\begin{subfigure}{0.48\textwidth}
   			\begin{tikzpicture}[scale=0.7, transform shape]
			% Parameters
			\def\squareSize{1cm} % Size of each square
			\def\spacing{0.2cm} % Space between squares
			\def\rows{2} % Number of rows
			\def\cols{5} % Number of columns
			\def\colsm{4}
			% Loop to draw the grid
			
			\foreach \i in {0,...,\rows} {
				\foreach \j in {0,...,2} {
					% Draw the square
					\draw[rounded corners=5pt, red] 
					(\j * \squareSize + \j * \spacing, -\i * \squareSize - \i * \spacing) 
					rectangle ++(\squareSize, \squareSize);}}		
			
			\foreach \i in {0,...,\rows} {
				\foreach \j in {3,...,5} {
					% Draw the square
					\draw[rounded corners=5pt, green] 
					(\j * \squareSize + \j * \spacing, -\i * \squareSize - \i * \spacing) 
					rectangle ++(\squareSize, \squareSize);}}	
			
			\foreach \i in {0,...,\rows} {
				\foreach \j in {6,...,8} {
					% Draw the square
					\draw[rounded corners=5pt, red] 
					(\j * \squareSize + \j * \spacing, -\i * \squareSize - \i * \spacing) 
					rectangle ++(\squareSize, \squareSize);}}

			\def\cornerRounding{5pt} % Rounding of corners
			
			%% at the top
			\foreach \i in {0,...,2} {
				\draw[rounded corners=\cornerRounding, red]  (\i * \squareSize+\i * \spacing,1.5*\squareSize+\spacing) -- ++(0, -0.5*\squareSize) -- ++(\squareSize, 0) -- ++(0, 0.5*\squareSize) ;
				\pgfmathsetmacro{\imone}{\i - 1}
				\pgfmathsetmacro{\ipl}{\i + 1}
				\draw[rounded corners=\cornerRounding,red] 
				(\ipl*\squareSize+\i*\spacing,-3*\squareSize-0.5*\spacing) -- ++(0, 0.5*\squareSize) -- ++(-\squareSize, 0) -- ++(0, -0.5*\squareSize) ;
			}
			
			%% at the top
			\foreach \i in {3,...,5} {
				\draw[rounded corners=\cornerRounding, green]  (\i * \squareSize+\i * \spacing,1.5*\squareSize+\spacing) -- ++(0, -0.5*\squareSize) -- ++(\squareSize, 0) -- ++(0, 0.5*\squareSize) ;
				\pgfmathsetmacro{\imone}{\i - 1}
				\pgfmathsetmacro{\ipl}{\i + 1}
				\draw[rounded corners=\cornerRounding,green] 
				(\ipl*\squareSize+\i*\spacing,-3*\squareSize-0.5*\spacing) -- ++(0, 0.5*\squareSize) -- ++(-\squareSize, 0) -- ++(0, -0.5*\squareSize) ;
			}
			
			%% at the top
			\foreach \i in {6,...,8} {
				\draw[rounded corners=\cornerRounding, red]  (\i * \squareSize+\i * \spacing,1.5*\squareSize+\spacing) -- ++(0, -0.5*\squareSize) -- ++(\squareSize, 0) -- ++(0, 0.5*\squareSize) ;
				\pgfmathsetmacro{\imone}{\i - 1}
				\pgfmathsetmacro{\ipl}{\i + 1}
				\draw[rounded corners=\cornerRounding,red] 
				(\ipl*\squareSize+\i*\spacing,-3*\squareSize-0.5*\spacing) -- ++(0, 0.5*\squareSize) -- ++(-\squareSize, 0) -- ++(0, -0.5*\squareSize) ;
			}

			%% at the right
			\draw[rounded corners=\cornerRounding, blue] 
			(9.5*\squareSize+9*\spacing,-2.5*\squareSize-.5*\spacing) -- ++(-0.5*\squareSize,0) -- ++(0,-0.5*\squareSize); 		
			\draw[rounded corners=\cornerRounding, blue] 
			(9.5*\squareSize+9*\spacing,1*\squareSize+1*\spacing) -- ++(-0.5*\squareSize,0) -- ++(0,0.5*\squareSize);

			\draw[rounded corners=\cornerRounding, blue] 
			(9.5*\squareSize+9*\spacing,-1.5*\squareSize+1.5*\spacing) -- ++(-0.5*\squareSize,0) -- ++(0,\squareSize) -- ++(0.5*\squareSize,0) ;
			\draw[rounded corners=\cornerRounding, blue] 
			(9.5*\squareSize+9*\spacing,-2.5*\squareSize+.5*\spacing) -- ++(-0.5*\squareSize,0) -- ++(0,\squareSize) -- ++(0.5*\squareSize,0) ;
			\draw[rounded corners=\cornerRounding, blue] 
			(9.5*\squareSize+9*\spacing,-0.5*\squareSize+2.5*\spacing) -- ++(-0.5*\squareSize,0) -- ++(0,\squareSize) -- ++(0.5*\squareSize,0) ;
			
			%% at the left
			\draw[rounded corners=\cornerRounding, blue] 
			(-1*\squareSize+1.5*\spacing,-2.5*\squareSize-0.5*\spacing) -- ++(0.5*\squareSize,0) -- ++(0,-0.5*\squareSize) ;	
			\draw[rounded corners=\cornerRounding, blue] 
			(-1*\squareSize+1.5*\spacing,1*\squareSize+1.*\spacing) -- ++(0.5*\squareSize,0) -- ++(0,0.5*\squareSize) ;		
			
			\draw[rounded corners=\cornerRounding, blue] 
			(-1*\squareSize+1.5*\spacing,-1.5*\squareSize+0.5*\spacing) -- ++(0.5*\squareSize,0) -- ++(0,-\squareSize) -- ++(-0.5*\squareSize,0) ;
			
			\draw[rounded corners=\cornerRounding, blue] 
			(-1*\squareSize+1.5*\spacing,-0.5*\squareSize+1.5*\spacing) -- ++(0.5*\squareSize,0) -- ++(0,-\squareSize) -- ++(-0.5*\squareSize,0) ;
			
			\draw[rounded corners=\cornerRounding, blue] 
			(-1*\squareSize+1.5*\spacing,.5*\squareSize+2.5*\spacing) -- ++(0.5*\squareSize,0) -- ++(0,-\squareSize) -- ++(-0.5*\squareSize,0) ;
			%% vertical lines
			\foreach \j in {-1,...,9} {
				\draw[gray!40,line width=1pt] (0.5*\squareSize+\j * \squareSize + \j * \spacing,0) -- ++(0,-\spacing);
				\draw[gray!40,line width=1pt] (0.5*\squareSize+\j * \squareSize + \j * \spacing,-\squareSize - \spacing) -- ++(0,-\spacing);
			}
			\foreach \j in {-1,...,9} {
				\draw[gray!40,line width=1pt] (0.5*\squareSize+\j * \squareSize + \j * \spacing,1*\squareSize +1*\spacing) -- ++(0,-\spacing);
				\draw[gray!40,line width=1pt] (0.5*\squareSize+\j * \squareSize + \j * \spacing,-2*\squareSize - 2*\spacing) -- ++(0,-\spacing);
			}
			%% horizontal lines
			\foreach \j in {-2,...,0} {
				\foreach \i in {0,...,9}{
					\draw[gray!40,line width=1pt] (\i*\squareSize+\i*\spacing,0.5*\squareSize+\j * \squareSize + \j * \spacing) -- ++(-\spacing,0);
				}
			}
			
			\foreach \j in {0,...,9} {
				\draw[gray!40,line width=1pt] (\j * \squareSize+\j * \spacing, -3.*\squareSize + 0.5*\spacing) -- ++(-\spacing,0);	
				\draw[gray!40,line width=1pt] (\j * \squareSize+\j * \spacing, 1.5*\squareSize + 0.5*\spacing) -- ++(-\spacing,0);	
			}

			\draw[fill=white, draw=white] (1.43*\squareSize+0*\spacing,-3.3) rectangle ++(0.25*\squareSize+1.5*\spacing,5);
			\draw[fill=white, draw=white] (4.43*\squareSize+3*\spacing,-3.3) rectangle ++(0.25*\squareSize+1.5*\spacing,5);
			\draw[fill=white, draw=white] (7.43*\squareSize+6*\spacing,-3.3) rectangle ++(0.25*\squareSize+1.5*\spacing,5);
			
			\node[blue] at (0*\squareSize-3*\spacing,0.5) {\huge$V$};
			\node[green] at (4*\squareSize+2*\spacing-1.1*\spacing,0.475) {\huge$V_m$};
			\node[blue] at (10*\squareSize+6*\spacing,0.5) {\huge$V$};
			\node[red] at (1*\squareSize-2*\spacing-0.1*\spacing,0.5) {\huge$U$};
			\node[red] at (7*\squareSize+3*\spacing+0.5*\spacing,0.5) {\huge$U$};
			
			\node[black] at (0*\squareSize-2*\spacing,-3.5) {\large$n=0$};
			\node[black] at (3*\squareSize-0.5*\spacing,-3.5) {\large$n=a$};
			\node[black] at (6*\squareSize+2.5*\spacing,-3.5) {\large$n=b$};
			\node[black] at (9*\squareSize+5.5*\spacing,-3.5) {\large$n=c$};
			
		\end{tikzpicture}
    \caption{Real-space lattice of the structure \JN{signifying the whole system in 2D, or  a slice from a three-dimensional lattice.}}
    \label{fig:onegate}\end{subfigure}\hfill
\begin{subfigure}{0.48\textwidth}
   \begin{tikzpicture}[scale=.7, transform shape,every node/.style={scale=.65}]
\draw [fill=blue] (-3,0.443934) rectangle (0,0.743934) node[pos=.5, text=white] {$V  - T_\mathbf{k}$};

\draw[fill=black] (0, -0.144165) rectangle (2, 0.0214847)  node[pos=.5, text=white] { $\approx 0  - T_\mathbf{k}/2$};
\draw[fill=black] (0, 0.888099) rectangle  (2, 1.02245) node[pos=.5, text=white] {\footnotesize{$\approx U  - T_\mathbf{k}/2$}};

\draw[fill=black] (4, -0.144165) rectangle (6, 0.0214847) node[pos=.5, text=white] {\footnotesize$\approx 0  - T_\mathbf{k}/2$};
\draw[fill=black] (4, 0.888099) rectangle  (6, 1.02245) node[pos=.5, text=white] {\footnotesize$\approx U  - T_\mathbf{k}/2$}; 

\draw[fill=blue] (6,0.443934) rectangle (9,0.743934)  node[pos=.5, text=white] {$V  - T_\mathbf{k}$};
\draw[fill=red] (2, -0.056066) rectangle  (4, 0.243934)  node[pos=.5, text=white] {$V_m  - T_\mathbf{k}$};

\draw[->] (3,0.243934) -- (3,.6);
\draw[->] (3,-0.056066) -- (3,.-.4);
\draw[->] (-3.5,-1) -- (10,-1) node[right] {$n$};
\draw[->] (-3.5,-1) -- (-3.5,1) node[above] {$E$};
\end{tikzpicture}

\caption{Energy spectrum. The gate potential $V_m$ is adjustable by a voltage.}
\label{fig:band_structure_one}\end{subfigure}
   \hfill
\caption{One semiconductor with a gate voltage $V_m$ between the same Mott insulating material $U$, everything is connected to a source and drain lead $V$.}
\label{fig:figures1}
\end{figure*}

Again, we use a plane wave ansatz for the different semiconducting and the Mott insulating regions, the source and drain $s_\pm$, the Mott insulating layers $r_\pm$ and the gate voltage region $m_\pm$. Hence, the \textit{ansatz} is given by
\begin{equation}
\label{eq:mottstructureansatz}
    p_n^0 = \begin{cases}
    s_+^n+R s_-^n & n \leq 0, \\
    A r_{+}^n+B r_{-}^n & \ 0<n \leq a, \\
    C m_+^n + D m_-^n & a < n \leq b, \\
    E r_{+}^n+ F r_{-}^n,  & b < n \leq c, \\
    \mathcal{T} s_+^n & c < n.
    \end{cases}
\end{equation}
The solutions in the various regions are \JV{in accordance to Eqs.~\ref{eq:ksemi} and \ref{Mott} and thus given by}
\begin{align}
\label{eq:ksemi}
s_\pm = e^{i \kappa_{\mathrm{semi}}n}, & \quad \cos\kappa_\mathrm{semi}=\frac{Z}{2T}\left[V-E-T_{\bf k}^\|\right],\\
m_\pm = e^{i \kappa_{\mathrm{m}}n}, & \quad \cos\kappa_\mathrm{m}=\frac{Z}{2T}\left[V_m-E-T_{\bf k}^\|\right], \\
r_\pm = e^{i \kappa_{\mathrm{Mott}}n}, & \quad\cos\kappa_\mathrm{Mott}=\frac{Z}{2T}\left[\frac{E(U-E)}{E-U/2}-T_{\bf k}^\|\right].
\end{align}
\JV{Due do the different structure the boundary conditions obtained from Eq.~\ref{difference} change (see App.~\ref{app:w1w2}).}
The transmission amplitude \JV{we find} reads
\begin{equation}
\begin{aligned}
    \mathcal{T}(E,k^\|)=& \frac{\left(m^2-1\right) \left(r^2-1\right)^2 \left(s^2-1\right) \left(-m^{a+b}\right) r^{a+b+c+1}}{s^{c+1} \big( W_1 + W_2 \big)}
       \end{aligned}
\end{equation}
with long expressions $W_1$ and $W_2$ given in App.~\ref{app:w1w2}.
The transmitted current is given by the absolute square $|\mathcal{T}(E,k^\|)|^2$ \cite{verlage2024quasi}, because the leads have the same effective wave number. \\

\subsection{Transmittance}
In the current context, we can discern three primary scenarios. Firstly, the potential at the source and drain terminals, denoted as $V$, may fall below the lower Hubbard band. Alternatively, it could lie within the region between the two bands. Lastly, it might exceed the upper Hubbard band. Additionally, there is a possibility that the potential is (partially) situated within one of the two Mott bands. In this case, the presence of such a potential configuration results in the emergence of a transmission channel.
\JVn{The discussion can be simplified substantially by using the particle-hole symmetry \cite{verlage2024quasi} in the Mott insulating layers. The particle $p_n^1$ and hole $p_n^0$ solutions are related via $(E-U)p_n^1=E p_n^0$ which shows the symmetry around $U/2$, the middle of the band gap. Therefore, it is enough to look at three different cases: $V$ below lower Hubbard band, $V$ inside lower Hubbard band and $V$ inside the Mott gap but below $U/2$. The other ones can be obtained by the symmetry operation of inversion at the $U/2$ energy with appropriately shifting the  on site potentials.}
\JN{In the following, we will use $T=0.3U$ as the hopping strength, measured in units of the Coulomb interaction, which is providing the energy scale. This yields an effective strength of $T/Z=0.075U$. This is in the same range as in other studies using $0.05 <T/U<0.2$ \cite{Okamoto2004,okamoto2007nonequilibrium,popovic2005wedge,okamoto2006lattice,leonov2015metal}. 
%Nevertheless, the qualitative results will not change with 
Moreover, the qualitative behavior turns out to be independent of 
the hopping strength, as long as the parameter choice stays within the strongly correlated Mott insulator regime. Therefore, we believe that the discussion below is %our results are 
transferable to other parameter combinations.}

\subsubsection{$V=-0.3U$ -- below lower Hubbard band}
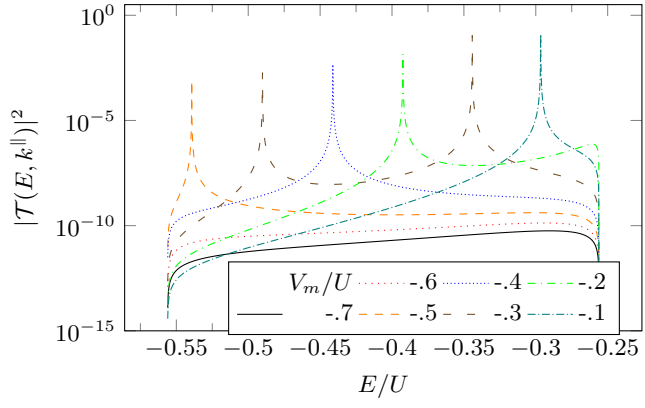
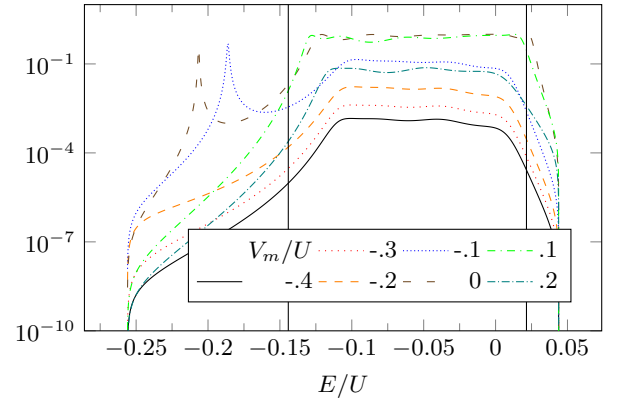
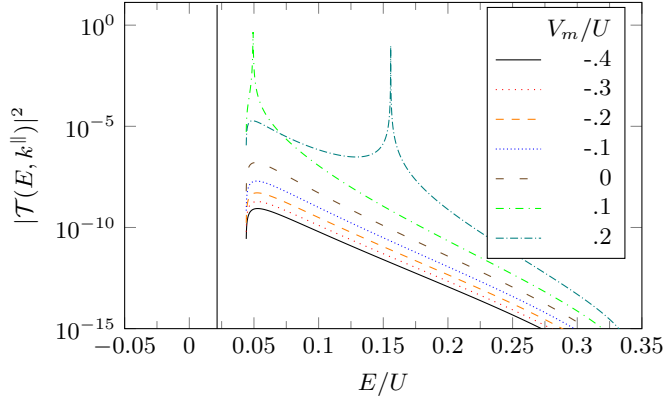
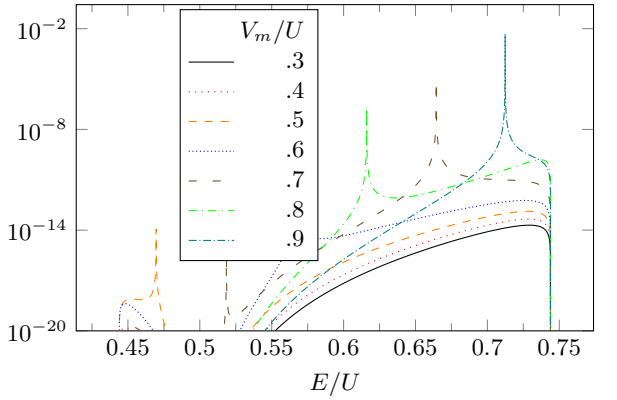
\begin{figure*}
\begin{subfigure}{0.47\textwidth}
  	\begin{tikzpicture}
		\begin{axis}[
            		ymin= 1e-15,
			xtick distance = .05,
			%ytick distance = 0.1,
			%grid = both,
			minor tick num = 1,
			major grid style = {lightgray},
			minor grid style = {lightgray!25},
			width = 1.\linewidth,
			height = 0.7\linewidth,
			xlabel = {$E/U$},
			ylabel ={$|\mathcal{T}(E,k^\|)|^2$},
			legend cell align = {right},
legend columns=2,transpose legend,
            legend style={at={(.2,.1)},anchor=west},
            ymode=log,
            cycle list={%
    {black},
    {red, dotted},
    {orange, dashed},
    {blue, densely dotted},
    {brown!60!black,loosely dashed},
    {green, dashdotted},
    {teal,densely dashdotted}
    }  ,
  ticklabel style={
	/pgf/number format/.cd,
	/pgf/number format/fixed,
%	use comma,% Komma als Dezimaltrenner
	1000 sep = {}% keine Tausendertrennung
},
scaled ticks=false   
                       			]
		 \addlegendimage{empty legend}
			\foreach \x in {1,2,3,4,5,6,7}
			\addplot table[x index=0, y index=\x] {vm0p3_onegate.dat};
				
		\draw[] (-0.144165,1e-10) -- ( -0.144165,10);			
				\draw[] (0.0214847,1e-10) -- ( 0.0214847,10);

			\legend{$V_m/U$,-.7, -.6, -.5, -.4, -.3, -.2, -.1, 0, .1, .2};
					
		\end{axis}

	\end{tikzpicture}
    \caption{(Color online)  With $V=-0.3U$ the energy range is below the lower Hubard band }
    \label{fig:vm0p3_one}
\end{subfigure}
\hfill
\begin{subfigure}{0.47\textwidth}
  	\begin{tikzpicture}
		\begin{axis}[
            		ymin= 1e-10,
			xtick distance = .05,
			%ytick distance = 0.1,
			%grid = both,
			minor tick num = 1,
			major grid style = {lightgray},
			minor grid style = {lightgray!25},
			width = 1.\linewidth,
			height = 0.7\linewidth,
			xlabel = {$E/U$},
			legend cell align = {right},
            legend style={at={(.2,.2)},anchor=west},
            ymode=log,
            legend columns=2,transpose legend,
             cycle list={%
    {black},
    {red, dotted},
    {orange, dashed},
    {blue, densely dotted},
    {brown!60!black,loosely dashed},
    {green, dashdotted},
    {teal,densely dashdotted}
    },
  ticklabel style={
	/pgf/number format/.cd,
	/pgf/number format/fixed,
%	use comma,% Komma als Dezimaltrenner
	1000 sep = {}% keine Tausendertrennung
},
scaled ticks=false    
           			]
		 \addlegendimage{empty legend}
			\foreach \x in {1,2,3,4,5,6,7}
			\addplot table[x index=0, y index=\x] {v0p0_onegate.dat};
				
		\draw[] (-0.144165,1e-10) -- ( -0.144165,10);			
				\draw[] (0.0214847,1e-10) -- ( 0.0214847,10);				
			
		\legend{$V_m/U$,-.4, -.3, -.2, -.1, 0, .1, .2, .3, .4, .5};
					
		\end{axis}

	\end{tikzpicture}
    \caption{(Color online)  With $V=0U$ the lower Hubbard band is fully covered.}
    \label{fig:v00_one}
    \end{subfigure}
\hfill
\begin{subfigure}{0.47\textwidth}    	\begin{tikzpicture}
		\begin{axis}[
			xmin = -0.05, xmax =0.35,
            		ymin= 1e-15,
			xtick distance = .05,
			%ytick distance = 0.1,
			%grid = both,
			minor tick num = 1,
			major grid style = {lightgray},
			minor grid style = {lightgray!25},
			width = 1.\linewidth,
			height = 0.7\linewidth,
			xlabel = {$E/U$},
						ylabel ={$|\mathcal{T}(E,k^\|)|^2$},
			legend cell align = {right},
            legend style={at={(.7,.6)},anchor=west},
            ymode=log,
             cycle list={%
    {black},
    {red, dotted},
    {orange, dashed},
    {blue, densely dotted},
    {brown!60!black,loosely dashed},
    {green, dashdotted},
    {teal,densely dashdotted}
    },
  ticklabel style={
	/pgf/number format/.cd,
	/pgf/number format/fixed,
%	use comma,% Komma als Dezimaltrenner
	1000 sep = {}% keine Tausendertrennung
},
scaled ticks=false   
           			]
 \addlegendimage{empty legend}
				
			\foreach \x in {1,2,3,4,5,6,7}
			\addplot table[x index=0, y index=\x] {v0p3_onegate.dat};
				
		\draw[] (-0.144165,1e-20) -- ( -0.144165,10);			
				\draw[] (0.0214847,1e-20) -- ( 0.0214847,10);				
			
		\legend{$V_m/U$,-.4, -.3, -.2, -.1, 0, .1, .2, .3, .4, .5};
					
		\end{axis}

	\end{tikzpicture}
    \caption{(Color online) The electron energy is in between the two Hubbard bands, below the point $E=U/2$.}
    \label{fig:v03_one}
\end{subfigure}
\hfill
   \begin{subfigure}{0.47\textwidth}     
             	\begin{tikzpicture}
		\begin{axis}[
            		ymin= 1e-20,
			xtick distance = .05,
			%ytick distance = 0.1,
			%grid = both,
			minor tick num = 1,
			major grid style = {lightgray},
			minor grid style = {lightgray!25},
			width = \linewidth,
			height = 0.7\linewidth,
			xlabel = {$E/U$},
			legend cell align = {right},
            legend style={at={(.2,.6)},anchor=west},
            ymode=log,
             cycle list={%
    {black},
    {red, dotted},
    {orange, dashed},
    {blue, densely dotted},
    {brown!60!black,loosely dashed},
    {green, dashdotted},
    {teal,densely dashdotted}
    },
  ticklabel style={
	/pgf/number format/.cd,
	/pgf/number format/fixed,
%	use comma,% Komma als Dezimaltrenner
	1000 sep = {}% keine Tausendertrennung
},
scaled ticks=false   
           			]
		 \addlegendimage{empty legend}
			\foreach \x in {4,5,6,7,8,9,10}
			\addplot table[x index=0, y index=\x] {v0p7_onegate.dat};
				
		\draw[] (-0.144165,1e-10) -- ( -0.144165,10);			
				\draw[] (0.0214847,1e-10) -- ( 0.0214847,10);

			\legend{$V_m/U$, .3, .4, .5, .6, .7, .8, .9};
					
		\end{axis}

	\end{tikzpicture}
    \caption{(Color online)  The electron energy is in between the two Hubbard bands, above the point $E=U/2$.}
    \label{fig:v07_one}
    \end{subfigure}
    \hfill
\caption{\JV{Transmission for $k^\|=\pi/4$ through the semiconductor-Mott heterostructure as a function of electron energy for different on-site potentials relative to the Hubbard bands.} The plot range always gives the band of the source/drain, the different colors depict different $V_m/U$. Vertical lines depict the Hubbard bands, if lying in the depicted range. The parameters are $T =0.3U$, $Z = 4$, $a = 2$, $ b = 5$, $c= 6$}
\label{fig:figures2}
\end{figure*}

Using $V=-0.3U$ and a specific parallel momentum value $k^\|$, the  transmittance exhibits a pronounced peak. This resonance energy \JV{shifts with the gate potential} $V_m$. It \JV{is} important to note that these resonant states do not manifest for all values of $V_m$. Instead, we observe them solely within a range of gate potentials $-0.4< V_m/U<-0.2$ (Fig.~\ref{fig:vm0p3_one}). \JVn{The resonances only appear if there is a band overlap between the leads and the gate, as they stem from constructive interference within the gate region.}

\subsubsection{$V=0U$ -- lower Hubbard band}
In the $V=0U$ case, 
the lead bands cover the lower Hubbard band fully covered, but reach down further in energy. By varying the
gate potential, $V_m$ we find peaks below the lower Hubbard band, see Fig.~\ref{fig:v00_one}.
Inside the lower Hubbard band there is a transmission channel with \JV{nearly constant} transmission for all gate potentials; they only \JV{differ in} the absolute value of the transmission. Above the band, transmission sharply drops off to zero {\JVn{(besides exponentially small tunneling currents)}}.

\subsubsection{$V=0.3U$ \& $V=0.7U$-- \JV{inside Mott gap}}
In \cite{verlage2024quasi}, it was demonstrated that \JV{there is exactly zero transmission through Mott layers in the middle of the Mott gap $E=U/2$ with an unpolarized spin background, which is the case here.} At this particular energy level, a perfect destructive interference between particle and hole currents occurs. Consequently, it is crucial to differentiate between situations where the energy is below or above this specific point.
In Fig.~\ref{fig:v03_one}, we observe the first scenario, where the energy is below the midpoint of the band gap. Similar to previous observations, a peaked transmission structure is obtained for specific values of the gate potential $V_m$, while for other values, the transmission remains nearly constant close to zero.

For the second scenario, where the energy is above the midpoint of the band gap, refer to Fig.~\ref{fig:v07_one}. \JVn{The symmetry between to two cases below and above the midpoint of the Mott gap can be seen.}
%In this case, the transmission pattern closely resembles that observed below the midpoint of the band gap but with a mirrored configuration. \JV{This is %part 
%PK ??
%a consequence of the partice-hole symmetry.} It is worth noting that the absolute value of the transmission in this scenario is reduced by five orders of magnitude compared to the former case, \JVn{because the inversio}.

%\subsubsection{$V=1.1U$ --  upper Hubbard band}
%Similarly, as previously discussed, there remains the possibility of achieving peaked transmission probabilities above the upper Hubbard band, specifically within the range of $0.9U<V_m<1.3U$. 
%Within the upper Hubbard band itself, the presence of a transmission channel is characterized by a nearly constant transmission rate for every possible gate potential, as illustrated in Fig.~\ref{fig:v11_one}.
%\subsubsection{$V=1.3U$ -- above upper Hubbard band}
%When the energy exceeds the upper Hubbard band, the transmission behavior becomes sensitive to the degree of band mismatch. For instance, when $V=1.5U$, the transmission drops significantly to the order of $10^{-10}$. However, when $V=1.3U$, a peaked transmission structure reemerges. This phenomenon is visually represented in Fig.~\ref{fig:v13_one}. It is important to note that, similar to previous observations, there exists a specific value of the gate potential $V_m$ where the transmission peaks vanish.

\subsection{Resonances}
\begin{figure}[t]
	\centering
	\begin{subfigure}{0.48\textwidth}
		\begin{tikzpicture}
		\begin{axis}[
			xmin = 0.4, xmax =0.743934,
			ymin= 1e-21,
			xtick distance = .05,
			%ytick distance = 0.1,
			%grid = both,
			minor tick num = 1,
			major grid style = {lightgray},
			minor grid style = {lightgray!25},
			width = \linewidth,
			height = 0.7\linewidth,
			xlabel = {$E/U$},
			ylabel ={$|\mathcal{T}(E,k^\|)|^2$},
			legend cell align = {right},
			legend style={at={(0,.91)},anchor=west},
			legend columns = -1,
			ymode=log,
			]
			\addlegendimage{empty legend}
			
			\addplot[black] table[x index=0, y index=1] {resonances_one.dat};
			\addplot[blue, dashed] table[x index=0, y index=2] {resonances_one.dat};
			\addplot[red, dotted] table[x index=0, y index=3] {resonances_one.dat};
			
			%				\draw[-,black] 	(0.437868,1e-25) -- (0.437868,1) node[midway] {$n=1$};
			%	\draw[-,black] 	( 0.5,1e-25) -- ( 0.5,1) node[midway] {$n=2$};
			\draw[-,black] 	(0.606066,1e-25) -- (0.606066,1)  node[midway] {$n=3$};
			
			\draw[-,blue, dashed] 	(0.443934,1e-25) -- (0.443934,1) node[midway] {$n=1$};
			\draw[-,blue, dashed] 	(0.55,1e-25) -- (0.55,1) node[midway] {$n=2$};
			\draw[-,blue,dashed] 	(0.656066,1e-25) -- (0.656066,1)  node[midway] {$n=3$};
			
			\draw[-,red, dotted] 	(0.543934,1e-25) -- (0.543934,1) node[near end] {$n=1$};
			\draw[-,red, dotted] 	(0.65,1e-25) -- (0.65,1) node[near end] {$n=2$};
			%	\draw[-,red,dotted] 	(0.756066,1e-25) -- (0.756066,1)  node[near end] {$n=3$};

			\legend{$V_m/U$,0.65,0.7,0.8};
			
		\end{axis}
		
	\end{tikzpicture}
	\caption{(Color online) Transmittance \JV{vs. predicted} resonance energies.}
	\label{fig:v07resonances_one}
	\end{subfigure}
	\hfill
	\begin{subfigure}{0.48\textwidth}
		\begin{tikzpicture}
		\begin{axis}[%
			domain=0:4,
			restrict y to domain=-1:1,
			xmin=-4,xmax=4,
			ymin=-0.04,ymax=0.04,
			xtick distance = 1,
			ytick distance=.02,
			%ytick distance = 0.1,
			%grid = both,
			minor tick num = 1,
			major grid style = {lightgray},
			minor grid style = {lightgray!25},
			width = \linewidth,
			height = 0.7\linewidth,
			xlabel = {$V_m/U$},
			ylabel ={$a(V_m)$},
			legend cell align = {right},
			legend style={at={(0,.8)},anchor=west},
			legend columns = 2,
			]
			\addlegendimage{empty legend}
			
			\addplot+[black,mark size=1.5pt, only marks,mark options={solid}] table[x index=0, y index=1] {tabvv01dvt.dat};
			\addplot+[red,mark size=1.5pt, only marks,mark options={solid}] table[x index=0, y index=1, mark=*] {tabvv005dvt.dat};
			\addplot+[blue,mark size=1.5pt, only marks,mark options={solid},mark=square] table[x index=0, y index=1] {tabvv00dvt.dat};
			
			\addplot+[orange,mark size=1.5pt, only marks,mark options={solid}, mark=triangle] table[x index=0, y index=1, only marks] {tabvvm005dvt.dat};
			\addplot+[mark options={solid},teal,mark size=1.5pt, only marks, mark=diamond] table[x index=0, y index=1] {tabvvm01dvt.dat};
			
			\legend{$\frac{V_m-V}{U}$,0.1,0.05,0,-0.05,-0.1};
			
			\draw[black] (-4,0)--(4,0);
			\draw[black, dashed] (0.5,-0.7)--(0.5,.7);
			
			\draw[black,dotted] (-0.144165,-.7)--(-0.144165,.7);
			\draw[black,dotted] (0.888099,-.7)--(0.888099,.7);
			\draw[black,dotted] (0.0214847,-.7)--(0.0214847,.7);
			\draw[black,dotted] (1.02245,-.7)--(1.02245,.7);

		\end{axis}
	\end{tikzpicture}
	\caption{(Color online) Renormalization parameter $a(V_m)$ of the gate voltage. The dashed line marks $U/2$, the dotted lines give the Mott bands.}
	\label{fig:renorm_para_a}
\end{subfigure}

	\caption{(a) gives the transmittance \JV{as function of energy for a lead potential }$V=0.7U$ together with the resonance energies predicted by Eq.~\ref{eq:res_en_one} as dashed vertical lines \JV{for different gate potentials $V_m$}. (b) gives the renormalization parameter $a(V_m)$ for different for various combinations of $V$ and $V_m$. The parameters are $T =0.3U$, $Z = 4$, $a = 2$, $ b = 5$, $c= 6$, $k^\| = \pi/4$.}
	\label{fig:figures3}
\end{figure}
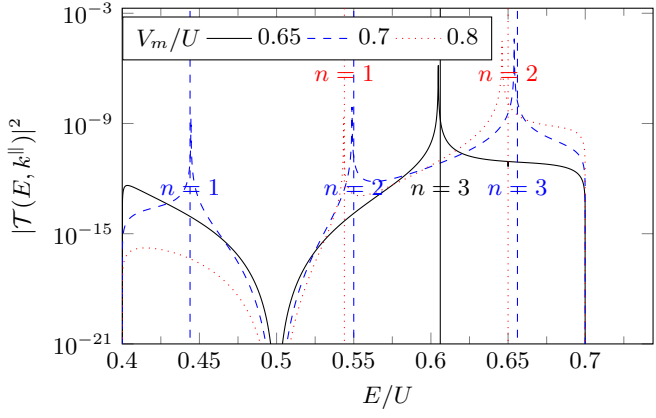
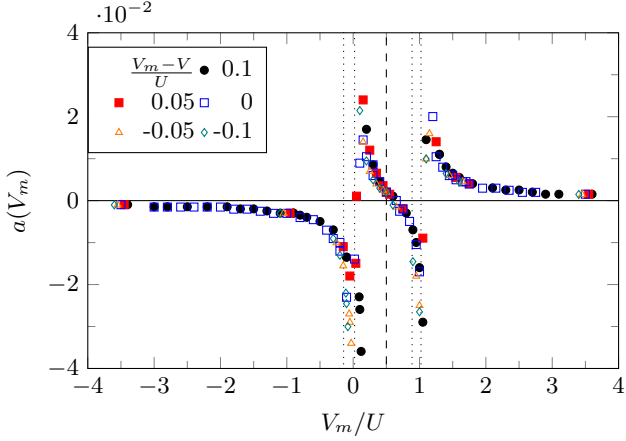

In the various scenarios discussed \JV{above}, we observe peaks or resonances in the transmittance \JVn{outside the Mott bands}. \JVn{Because of this,} these resonances \JVn{can only} arise from constructive interference within the gate and should fulfill $ \kappa_m d = n \pi$.
Here,  $\kappa_m$ is the wave number, $d$ is the thickness of the gate layer, and $n\in \mathbb{Z}$  is an integer.
These conditions lead to resonance energies $ E_\mathrm{res} $ given by:
\begin{equation}
\label{eq:res_en_one}
E_\mathrm{res}=\frac{-2T \cos(k^\|)+V_m Z-2 T \cos(\frac{\pi n}{d})}{Z}.
\end{equation}
However, when comparing these predicted resonance energies to the actual ones, $E_\mathrm{res}^\mathrm{real}$ obtained from $|\mathcal{T}|^2 $ (Fig.~\ref{fig:v07resonances_one}), we notice discrepancies. Assuming the correctness of the resonance energy formula, we introduce a renormalized gate voltage $ V_m^r = V_m + a(V_m) $. Here, if $ a(V_m) $ is negative, it signifies a downward shift in the gate voltage.
In Fig.~\ref{fig:renorm_para_a}, we present  $a(V_m)$ for different cases where $|V_m - V| \leq T $, ensuring an overlap between the source/drain and gate bands to maintain resonances. Several observations can be made:

\JV{a) If the energy falls inside one of the Mott bands, it opens up a transmission channel throughout the structure. Thus the effective width of the region wherein constructive interference happens is larger than the gate region.} \JVn{This renders the resonance energy formula Eq.~\ref{eq:res_en_one} wrong, because there might be modes fitting in more than one region now.}
%The thickness of the layer $d$ here is not just given by the gate voltage region, but also by the Mott insulating ones. Because of this, \JV{the $d$ in the formula used is the size of the entire three layers. Because this is not done, }
The renormalization $a(V_m)$ compensates for this effect.

\JV{b)} Outside the Mott bands, where the thickness of the layer is solely given by the gate region, the 
renormalization pushes the 
\JV{effective} gate potential $V_m$ away from the bands. Below the lower band, it is pushed towards smaller values with $a(V_m)<0$, while above the upper band, it is pushed towards larger values with $a(V_m)>0$.

\JV{c)} Between the two Mott bands, a change in sign occurs at $\approx U/2$ (the shift is due to $k^\|\neq 0$). Below this point, $a(V_m)$ is positive, while above it, $a(V_m)$ is negative. \JV{Inside the Mott gap the effective gate potential is pushed towards the middle of the gap.}

Overall, these trends are consistent across different $V_m - V$ cases, with small deviations attributable to numerical considerations. Thus, there is no dependency on the specific distance $V_m - V$,  and only the relative position of the gate voltage $V_m$ with respect to the Mott bands is of significance. Far away from the bands the renormalization goes to zero $a(V_m \gg U) = a(V_m \ll 0) \to 0$.

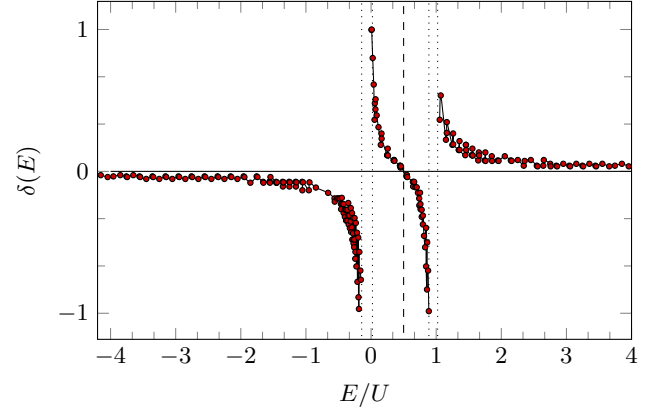
\begin{figure}
	\begin{tikzpicture}
		\begin{axis}[%
			no markers, smooth,
			domain=0:4,
			restrict y to domain=-1:1,
			xmin=-4.2,xmax=4,
			xtick distance = 1,
		%	ytick distance=.02,
			%ytick distance = 0.1,
			%grid = both,
			minor tick num = 2,
			major grid style = {lightgray},
			minor grid style = {lightgray!25},
			width = \linewidth,
			height = 0.7\linewidth,
			xlabel = {$E/U$},
			ylabel ={$\delta(E)$},
			legend cell align = {right},
			legend style={at={(0,.8)},anchor=west},
			legend columns = 2,
			]
			\addlegendimage{empty legend}
			
			\addplot+[black] table[x index=0, y index=1] {deltaphasescattlen.dat};
			\addplot+[black,mark size=1pt, only marks] table[x index=0, y index=1] {deltaphasescattlen.dat};

			\draw[black] (-4,0)--(4,0);
			\draw[black, dashed] (0.5,-1.7)--(0.5,1.7);
			
			\draw[black,dotted] (-0.144165,-1.7)--(-0.144165,1.7);
			\draw[black,dotted] (0.888099,-1.7)--(0.888099,1.7);
			\draw[black,dotted] (0.0214847,-1.7)--(0.0214847,1.7);
			\draw[black,dotted] (1.02245,-1.7)--(1.02245,1.7);

		\end{axis}
	\end{tikzpicture}
	\caption{(Color online) Phase shift $\delta(E)$ for $V=V_m$. The dashed line marks $U/2$, the dotted lines give the Mott bands.}
	\label{fig:deltaphase}
\end{figure}

\begin{figure}
    \centering
    \begin{tikzpicture}[scale=.8]
   \pgfplotsset{
        scale only axis,
     %   scaled x ticks=base 10:3,
     %   xmin=0, xmax=0.06,
        y axis style/.style={
            yticklabel style=#1,
            ylabel style=#1,
            y axis line style=#1,
            ytick style=#1
       }
    }    
    \begin{axis}[
    width = \linewidth,
			height = 0.7\linewidth,
      axis y line*=left,
      y axis style=blue!75!black,
      ymin=-2.1, ymax=1.5,
      xlabel=$\kappa$,
      	xtick distance = .2,
      ylabel={$\delta(\kappa)$},
     % ymode=log,
        legend style={at={(0.05,.8)},anchor=west},
        grid=both
    ]
    \addplot+[blue, mark options={solid},mark size=1pt, only marks, mark=square] table[skip first n=0,x index=0, y index=1] {dka.dat};

    \end{axis}
    
    \begin{axis}[
    width = \linewidth,
			height = 0.7\linewidth,
      axis y line*=right,
      axis x line=none,
      ymin=-2, ymax=2,
      ylabel=$a_s(\kappa)$,
      y axis style=red!75!black,
       legend style={at={(0.7,.3)},anchor=west}
    ]
\addplot+[red,mark=*, mark options={solid},only marks, mark size=1pt] table[skip first n=0,x index=0, y index=2] {dka.dat};

    \end{axis}
    
    \end{tikzpicture}
    \caption{(Color online) Phase shift $\delta(\kappa)$ \JV{(blue squares, left axis)} and scattering length $a_s(\kappa)$ \JV{(red circles, right axis)} \JV{for $V=V_m$.} \JV{The oscillations are an artifact of the energy resolution used in the analysis.} }
    \label{fig:dka}
\end{figure}
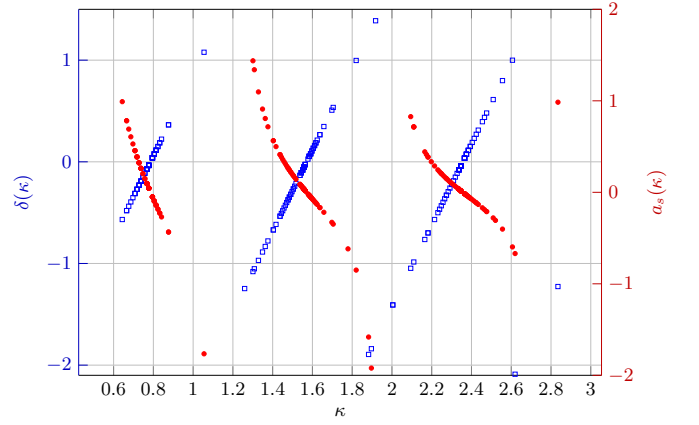

\subsection{Scattering Phase}
\JV{In order to obtain the renormalized gate potential to explain the shift between the predicted and real resonances we 
need to assume the correctness 
assumed $\kappa_m d=n\pi$ to be valid. This is 
%not necessarily true, within a 
different for the phase accumulation model for quantum wells \cite{shikin2001phase,smith1985phase,danese2002phase}: there, the scattering phase $\delta(E)$ }
%The presence of the scattering phase 
modifies the resonance condition to 
\begin{equation}
\kappa_m d = n \pi + \delta(E).
\end{equation}
This scattering phase is energy-dependent and leads to resonance energies described by 
\begin{equation}
\label{eq:res_en_one_delta}
E_\mathrm{res}^\delta = \frac{-2T \cos(k^\|) + V_m Z - 2T \cos\left(\frac{\pi n + \delta(E)}{d}\right)}{Z}.
\end{equation}
From these resonance energies, we can deduce the behavior of the scattering phase, $\delta(E)$, as depicted in Figure~\ref{fig:deltaphase}. \JV{The energy range is covered by tuning the lead potential $V$.} Notably, like the renormalization of the potential, the scattering phase undergoes a sign change at the Mott bands and at $U/2$. As one moves far away from the Mott bands, the scattering phase tends to zero.
%{\color{blue}{Why? A negativ scattering phase means a pulling into the potential, a positive one means a push away}}
\JV{Functionally, the energy dependent scattering phase $\delta(E)$ depends on the band width contribution $W=\frac{2T}{Z}$ from the hopping orthogonal to the interfaces. With this, it reads
\begin{equation}
    \delta(E)= \begin{cases}
        \frac{W}{2}\frac{1}{E+\frac{W}{2}} & E< \text{lower Hubbard band}, \\
        -W \tan\left(\pi \left[\frac{E}{U}-\frac{U}{2} \right] \right) & \text{between Hubbard bands}, \\
        \frac{W}{2}\frac{1}{E-U+\frac{W}{2}} & E> \text{upper Hubbard band.} 
    \end{cases}
\end{equation}}
As expected, the maximum of the derivative of the scattering phase coincides with the Mott bands. This observation is in \JV{line} with the fact that this quantity essentially describes the density of states \cite{PhysRevLett.114.033901}.

From the scattering phase and the wave vector $\kappa$, we can also compute the scattering length $a_s(\kappa)=\frac{-\tan(\delta(\kappa))}{\kappa}$.
Fig.~\ref{fig:dka} \JV{shows} the scattering phase $\delta(\kappa)$ together with $a_s(\kappa)$. \JV{There are several branches of the momentum dependent scattering phase $\delta(k)$ (blue squares) that show a linear dependence on the wave number. For every branch the scattering length thus is given by $a_s(\kappa) \propto \tan(\mathrm{const}_1 \kappa+\mathrm{const}_2)/\kappa$ (red circles).}

\subsection{Parallel Momentum Integrated Transmission}
In practical real-world setups, controlling the value of $k^\|$ for electrons can be challenging. \JV{The electrons occupy all possible parallel momentum values, and in order to incorporate the fact that $k^\|$ is not resolved,} we will consider transmission probabilities that are \JV{averaged} over $k^\|$, denoted as $\mathcal{T}(E)$:
\begin{equation}
 \mathcal{T}(E) = \frac{1}{N} \int \mathcal{T}(E, k^\|) \, dk^\| .
\end{equation}
The resulting  transmittance represents a superposition of individual transmission probabilities, leading to a superposition of peaks in the transmission spectrum. Tuning the gate voltage \(V_m\) shifts the region of appreciable transmission to either lower or higher energies.

\begin{figure}
	\centering
	\begin{subfigure}{.45\textwidth}
		\begin{tikzpicture}
		\begin{axis}[
			%xmin = 0.4, xmax =0.743934,
			%ymin = -2, ymax = 2,
			%ymin= 1e-21,
			xtick distance = .1,
			%ytick distance = 0.1,
			%grid = both,
			minor tick num = 1,
			major grid style = {lightgray},
			minor grid style = {lightgray!25},
			width = \textwidth,
			height = 0.7\linewidth,
			xlabel = {$E/U$},
			ylabel ={$|\mathcal{T}(E)|^2$},
			legend cell align = {right},
			legend style={at={(.4,.3)},anchor=west},
			ymode=log,
			cycle list={%
				{black},
				{red, dashed},
				{orange, dashed},
				{blue, dotted},
				{brown!60!black,loosely dashed},
				{green, dashdotted},
				{teal,densely dashdotted}
			}
			]
			\addlegendimage{empty legend}
		%	\foreach \x in {1,5}
		%	\addplot table[skip first n=1,x index=0, y index=\x] {tab_integratedkp_v07.dat};
			
			\foreach \x in {1,2}
			\addplot table[skip first n=1,x index=0, y index=\x] {tab_integratedkp_v07_betterresolution_2.dat};

			\legend{$V_m/U$, 0.5, 0.9};
			
		\end{axis}
		
	\end{tikzpicture}
    \vspace*{-5mm}
	\caption{(Color online) $V=0.7U$}
	\label{fig:v07_integratedkp}
	\end{subfigure}
	%\hfill
    
	\begin{subfigure}{.45\textwidth}
        \vspace{5mm}
		\begin{tikzpicture}
		\begin{axis}[
			ymin= 1e-8,
			xtick distance = .1,
			%ytick distance = 0.1,
			%grid = both,
			minor tick num = 1,
			major grid style = {lightgray},
			minor grid style = {lightgray!25},
			width = \textwidth,
			height = 0.7\linewidth,
			xlabel = {$E/U$},
			ylabel ={$|\mathcal{T}(E)|^2$},
			legend cell align = {right},
			legend style={at={(0.55,.6)},anchor=west},
%			legend columns = -1,
		%	ymode=log,
			cycle list={%
				{black},
				{red, dashdotted},
				{orange, dashed},
				{blue, dotted},
				{brown!60!black,loosely dashed},
				{green, dashdotted},
				{teal,densely dashdotted}
			}
			]
			\addlegendimage{empty legend}
			\foreach \x in {2,3,5}
			\addplot table[skip first n=1,x index=0, y index=\x] {tab_integratedkp_v11.dat};
			
			\draw[-](0,1)--(2,1);	
			\draw[dashed] (0.872015,1e-20) -- (0.872015,10);
			\draw[dashed] (1.08059,1e-20) -- (1.08059,10);
			\legend{$V_m/U$,   1.2,1.1, 0.9};
			
			% " 1.3", "1.2", "1.1", "1.0", "0.9", "0.8"	
		\end{axis}
		
	\end{tikzpicture}
     \vspace*{-3mm}
	\caption{(Color online) $V=1.1U$}
	\label{fig:v11_integratedkp}
\end{subfigure}

	\begin{subfigure}{.45\textwidth}
    \vspace{5mm}
		\begin{tikzpicture}
		\begin{axis}[
			ymin= 1e-9,
			xtick distance = .1,
			%ytick distance = 0.1,
			%grid = both,
			minor tick num = 1,
			major grid style = {lightgray},
			minor grid style = {lightgray!25},
			width = \textwidth,
			height = 0.7\linewidth,
			xlabel = {$E/U$},
			ylabel ={$|\mathcal{T}(E)|^2$},
			legend cell align = {right},
			legend style={at={(0.4,.7)},anchor=west},
		%	legend columns = -1,
		%	ymode=log,
			cycle list={%
				{black},
				{red, dashdotted},
				{blue, dashed},
				{red, dotted},
				{brown!60!black,loosely dashed},
				{green, dashdotted},
				{teal,densely dashdotted}
			}
			]
			\addlegendimage{empty legend}
			\foreach \x in {1,3}
			\addplot table[skip first n=1,x index=0, y index=\x] {tab_integratedkp_v04_betterresolution.dat};
			
			\legend{$V_m/U$, 0.2 ,0.4};
			
			% ".2 "," .3 "," .4 "," .5 "," 0.6 "," 0.7 "	
		\end{axis}
		
	\end{tikzpicture}
     \vspace*{-3mm}
	\caption{(Color online) $V=0.4U$}
	\label{fig:v04_integratedkp}
	\end{subfigure}
	
	\caption{Parallel momentum \JV{averaged}  transmittance for different source/drain potentials $V$ \JV{as a function of electron energy}. Dashed lines mark the \JV{edges of the upper}  Hubbard band. The parameters are $T =0.3U$, $Z = 4$, $a = 2$, $ b = 5$, $c= 6$.}
	\label{fig:figures4}
\end{figure}
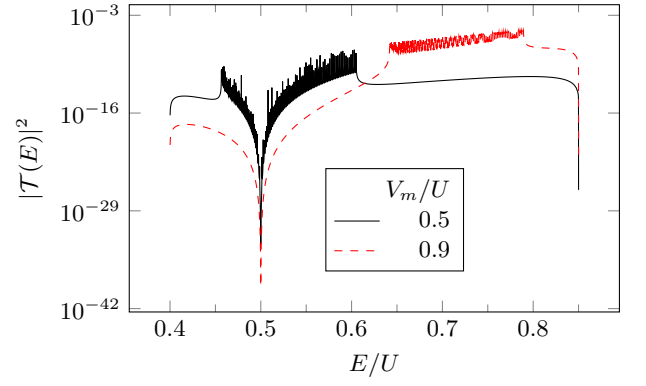
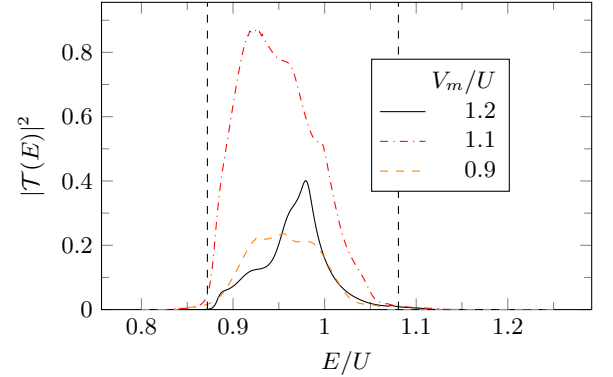
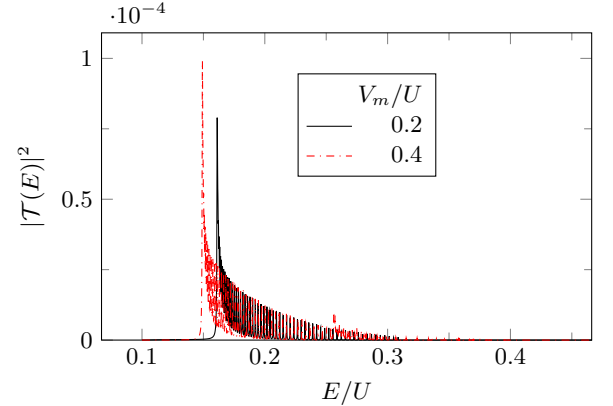

For example, when $V_m=0.9U$ (Fig.~\ref{fig:v07_integratedkp}), we observe a shift towards higher energies in the transmission spectrum for $V=0.7U$. Conversely, when $V_m$ is adjusted to $0.5U$, the transmission spectrum shifts towards lower energies. Also, the magnitude of the transmission goes down. Away from the occuring maximum transmittance, there is an exponential drop of.

Additionally, the higher peak in the transmission spectrum can also be shifted by varying the gate voltage, as demonstrated in Fig.~\ref{fig:v04_integratedkp}. In this case, transitioning from $V_m=0.2U$ to $0.4U$ causes the highest transmission peak to shift \JV{followed by} an exponential \JV{tail}, here to higher energies.
This exponential decay acts as the envelope of transmission peaks.

However, it is crucial to note that these observations primarily apply outside the Mott bands. Inside the Mott bands, changing the gate voltage tends to result in relatively flat transmission curves, as depicted in Fig.~\ref{fig:v11_integratedkp}. The alteration in gate voltage primarily affects the overall magnitude of transmission and the offset from zero outside the Mott bands. \JN{Changing the layer thickness by changing $a$, $b$ and $c$ does not qualitatively change the results. Transmission channels stay open, only evanescent waves yield a lower transmission with increasing the layer thickness.}

\subsection{Applying an Overall Voltage}
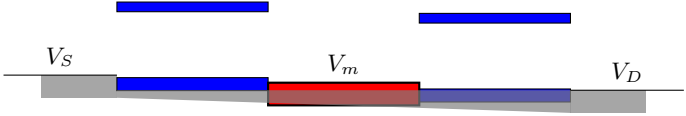
\begin{figure}
\begin{center}
   \begin{tikzpicture}[scale=1, transform shape,every node/.style={scale=1}]

	\fill [fill=gray, opacity=0.7] (-1,0.4) rectangle (0,0.7);
	
	\draw[fill=blue] (0, 0.494651) rectangle (2, 0.666667);
	\draw[fill=blue] (0, 1.53868) rectangle  (2, 1.66667);
	
	\draw[fill=blue] (4, 0.344651) rectangle (6, 0.516667);
	\draw[fill=blue] (4, 1.38868) rectangle  (6, 1.51667);
	
	\fill[fill=gray, opacity=0.7] (6,0.2) rectangle (7,0.5) ;
	\draw[fill=red,thick] (2, 0.3) rectangle  (4, 0.6) ;

\fill [fill=gray, opacity=0.7] (6,.5) -- (0,.5) -- (0,.4) -- (6,.2) -- cycle;
	
	\draw (0,0.7) -- (-1.5,0.7) node[above, midway] {$V_S$};
		\draw (6,0.5) -- (7.5,0.5) node[above, midway] {$V_D$};
				\draw (2,0.6) -- (4,0.6) node[above, midway] {$V_m$};

\end{tikzpicture}
\caption{\enquote{Band structure} of the system with applied voltage, the drain $V_D$ is lower than the source $V_S$, the gate potential $V_m$ is still adjustable. The applied voltage also shifts the Mott bands. Here, a linear behavior is assumed while inside the distinctive regions the potential is taken to be constant.}
\label{fig:system_with_voltage}
\end{center}
\end{figure}

Finally, we introduce a global voltage to the entire system. As depicted in Figure~\ref{fig:system_with_voltage} \JV{there is a potential difference between the source $V_S$ and the drain $V_D$ creating the driving voltage. This alters the on-site potential in all the layers, also affecting the Mott layers. Fig.~\ref{fig:system_with_voltage} shows a small applied voltage shifting the right Mott layer 
%more compared 
%PK 
relative to the left one. Inside the layers, for simplicity, the potential is assumed constant. For small fractions of voltage over system size this seems reasonable. With more sophisticated methods to solve all the boundary equations (e.g. transfer matrices) this can easily be refined.} 
\JVn{Consequently, we find the transmittance $T(E)$ as}
%Consequently, we can incorporate the resulting  transmittance into the Landauer-Büttiker formalism \cite{PhysRevB.31.6207,5392683} to compute the net current in this junction \cite{Wilhelm2013A}:
%
%\begin{equation}
%I(E) = \frac{e^2}{h}T(E) \big( f_D(E-V_D) - f_S(E-V_S) \big)
%\end{equation}
%
%In this equation, the Fermi-Dirac distribution function is denoted as $f_x(E) = \left(1 + \exp(\beta E)\right)^{-1}$, with an inverse temperature $\beta \in [0,\infty)$ for both the source $S$ and the drain $D$. The  transmittance, $T(E)$, is described as follows \cite{verlage2024quasi}:
%
\begin{equation}
T(E) = |\mathcal{T}(E)|^2 \frac{\sin(\kappa_D)}{\sin(\kappa_S)}.
\end{equation}
\JV{The additional factor $\frac{\sin(\kappa_D)}{\sin(\kappa_S)}$ comes because the effective wave numbers and thus the group velocities are not the same within the source and the drain.}

\begin{figure}
    \centering
    	\begin{tikzpicture}
		\begin{axis}[
			ymin= 1e-4,
			xtick distance = .1,
			%ytick distance = 0.1,
			%grid = both,
			minor tick num = 1,
			major grid style = {lightgray},
			minor grid style = {lightgray!25},
			width = 0.5\textwidth,
			height = 0.7\linewidth,
			xlabel = {$E/U$},
			ylabel ={$T(E)$},
			legend cell align = {right},
			legend style={at={(0.75,.8)},anchor=west},
			ymode=log,
			cycle list={%
				{black},
				{red, dashdotted},
				{blue, dashed},
				{red, dotted},
				{brown!60!black,loosely dashed},
				{green, dashdotted},
				{teal,densely dashdotted}
			}
			]
			\addlegendimage{empty legend}
			\foreach \x in {1,2,3}
			\addplot table[skip first n=0,x index=0, y index=\x] {dataTE_ohnefermi.txt};
			
			\legend{$V_m/U$, 0.9,1.1,1.3};
			
			% ".2 "," .3 "," .4 "," .5 "," 0.6 "," 0.7 "	
		\end{axis}
		
	\end{tikzpicture}
    \caption{\JV{Transmission through the semiconductor-Mott heterostructure with an applied voltage for a source potential $V_S=1.1U$ and a drain potential $V_D=0.7U$} as a function of energy for different gate potentials. The parameters are $T =0.3U$, $Z = 4$, $a = 2$, $ b = 5$, $c= 6$.}
    \label{fig:enter-label}
\end{figure}
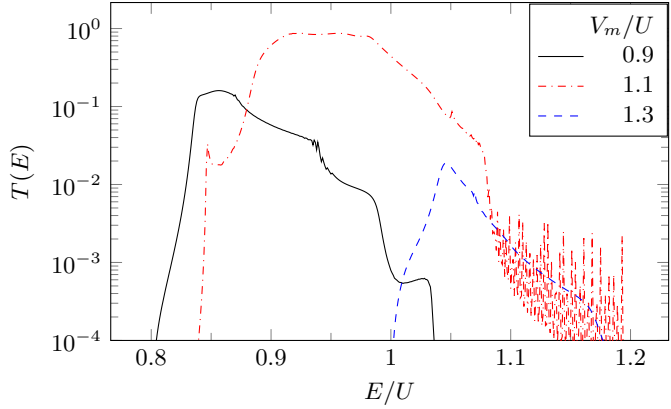

\JV{Fig.~\ref{fig:enter-label} shows the transmission for some exemplary values of the source, drain and gate potential.}
\JV{The gate voltage can be used to tune the energy range of the transmission. This is true within the energy range of the Mott bands and outside.}

\JN{Even though our analysis was done at zero temperature, the thermal energy of the quasi-particles $k_B T_{\mathrm{qp}}$ 
%PK viele Experimente werden gar nicht bei Raumtemperature gemacht
%, which is $25 \, \mathrm{meV}$ at room temperature, 
is much smaller than the Mott gap, $k_B T_{\mathrm{qp}} \ll U$ for typical materials. The quasi-particle structure remains valid even in the presence of scattering~\cite{queisser2019boltzmann}, and therefore equilibration may be described both inside the Mott bands, as well as at the source and drain, by applying a Fermi distribution $f(T_{\mathrm{qp}})$. In the end, 
%we would need to convolute 
finite-temperature results may be obtained by a convolution of the transmission $T(E)$ with $f(T_{\mathrm{qp}})$.
%this Fermi distribution.
}

\section{Conclusions}
%%%%%%%%%%%%%%%%%%%%%%%%%%%%%%%%%%%%%%%%%%%%%%%%%%%%%%%%%%%%%%%%%%%%%%%%%%%%%%%%%%%
%%%%%%%%%%%%%%%%%%%%%%%%%%%%%%%%%%%%%%%%%%%%%%%%%%%%%%%%%%%%%%%%%%%%%%%%%%%%%%%%%%%
\JV{In summary, we studied different types of heterostructures built by Mott insulators with on-site Coulomb repulsion $U$ and semiconductors on-site potential $V$. The hierarchy-of-correlations approach allows us to treat the strongly interacting Mott insulators and the weakly interacting semiconductors within the same framework. Using this formal expansion into the inverse coordination number $1/Z$, we can assign 
%obtain 
effective wave vectors to the quasi-particles allowing us to calculate the transmission probability through various types of structures with and without an applied voltage.}
%transmission 
%quantum conductance through these structures via the Landauer-Büttiker formula after having obtained the transmission probability.}

For a a semiconductor structure, our approach reproduces the well-known result of transmission through minibands. 

\JV{The Mott insulator-semiconductor heterostructure, on the other hand, allows us to make use of the strong suppression of the quasi-particle current due to destructive interference of the particle and hole channels in the middle of the Mott insulator band gap. Within a phase accumulation model for quantum wells we determine analytical expressions for the 
%scattering phase, which depends 
resonance energies which depend on the energy relative to the Mott bands.}

\begin{comment}In this new configuration, semiconducting leads, serving as source and drain terminals, encase a Mott insulating material characterized by an on-site repulsion parameter denoted as $U$. Within this insulating region, there exists another semiconductor segment controlled by a gate voltage $V_m$.
Again, we calculate the  transmittance for electrons depending on the gate voltage and find for certain paramater combinations resonances in the transmission.
From these, we deduct that the gate potential gets renormalized by the Mott bands; effectively it feels a repulsion. Lastly, we caluclated the parallel momentum integrated transmission after appyling an overall voltage to plug this into the Landauer-Büttiker formalism. 
\end{comment}
Looking at the transmission curves integrated over parallel momentum, we find a skewness in the transmission 
function around its center that could have possible applications in thermoelectric devices. 
\JV{The energies for which transmission is found can be tuned by the gate.} 
In this setting, the skewness of the transport distribution function is desirable as it helps to increase the Seebeck coefficient~\cite{zheng2022asymmetrical}. \JV{This feature distinguishes heterostructures with a Mott insulator, i.e., with strongly interacting electrons, from a pure semiconductor structure.} 

%PK Brauchen wir diesen Abschnitt, um die Fragen der Referees zu beantworten?
%PK Wenn nein, würde ich ihn lieber herauslassen,
%PK da wir durch hinzufügen von neuen gedanken wieder neue Diskussionspunkte reinbringen (was ich lieber vermeiden würde).
%\JN{As mentioned, the hierarchy of correlations provides an iterative way to include higher-order effects. Since the translation invariance is broken in heterostructures, the back-reaction from the two-point correlations $\hat\rho_{\mu\nu}^{\rm corr}$ onto the mean-field background might lead to space charge layers, especially at finite temperatures (which is also tractable within the hierarchy \cite{queisser2023hierarchy}). Furthermore, higher-order correlations mediate effective interactions between the quasi-particles in the Mott insulator, giving rise to an effective Boltzmann equation \cite{queisser2019boltzmann}. Therefore, the quasi-particles acquire a finite lifetime, whose inverse scales with the density. For small densities, as investigated here, these effects are negligible.}

%%%%%%%%%%%%%%%%%%%%%%%%%%%%%%%%%%%%%%%%%%%%%%%%%%%%%%%%%%%%%%%%%%%%%%%%%%%%%%%%%%%
%%%%%%%%%%%%%%%%%%%%%%%%%%%%%%%%%%%%%%%%%%%%%%%%%%%%%%%%%%%%%%%%%%%%%%%%%%%%%%%%%%%
\section{Acknowledgement}
%%%%%%%%%%%%%%%%%%%%%%%%%%%%%%%%%%%%%%%%%%%%%%%%%%%%%%%%%%%%%%%%%%%%%%%%%%%%%%%%%%%
%%%%%%%%%%%%%%%%%%%%%%%%%%%%%%%%%%%%%%%%%%%%%%%%%%%%%%%%%%%%%%%%%%%%%%%%%%%%%%%%%%%
The authors thank F. Queisser and R. Schützhold for fruitful discussions and valuable feedback on the manuscript. Funded by the Deutsche Forschungsgemeinschaft (DFG, German Research Foundation) – Project-ID 278162697– SFB 1242.

\section{Appendix}
\subsection{Semiconductor Structure}
\label{app:semis}
From Eq.~\ref{difference} we can deduct two boundary equations per interface \JV{for the structure shown in Fig.~\ref{fig:system}. One for the lattice site left of the interface and one to the right of it. In this case the eight equations read (after some algebra)}
\begin{align}
    s_{1+}B+\frac{A}{s_{1+}}&=\left(\frac{1}{s_+}+Rs_+ \right) \\
    1+R&=(A+B) \\
    A s_{1+}^{a+1}+\frac{B}{s_{1+}^{a+1}}&= \left(C s_+^{a+1}+\frac{D}{s_+^{a+1}} \right) \\
    C s_+^a+\frac{D}{s_+^a}&=\left(A s_{1+}^a+\frac{B}{s_{1+}^a} \right) \\
   C s_+^{b+1}+\frac{D}{s_+^{b+1}}&= \left(s r_{2+}^{b+1} + \frac{F}{s_{2+}^{b+1}}\right) \\
    E s_{2+}^b +\frac{F}{s_{2+}^b}&= \left( C s_+^b+\frac{D}{s_+^b} \right) \\
    E s_{2+}^{c+1}+\frac{F}{s_{2+}^{c+1}}&= \mathcal{T} s_+^{c+1}\\
    \mathcal{T} s_+^c &= \left( E s_{2+}^c +\frac{F}{s_{2+}^c} \right).
\end{align}
\JV{Inserting the \textit{ansatz} \ref{eq:ansatzsemistructure} into this we can solve for the transmission coefficient behind the interface as}

\begin{equation}
\label{eq:trans}
    \mathcal{T}= \frac{-s_{1+}^{a+1}\left(s_{1+}^2-1\right) \left(s_{2+}^2-1\right) \left(s_+^2-1\right)^2  s_+^{a+b} s_{2+}^{b+c}}{s_+^{1 + c}\left( M_1 +M_2+M_3+M_4+M_5+M_6 \right)}
\end{equation}
with the six terms 
\begin{equation}
    \begin{aligned}
        M_1 =& s_{2+}^{2 b} \big(s_+^{2 a} \left(s_{1+}^{2 a+2} (s_{1+}-s_+)^2-(s_{1+} s_+-1)^2\right) \\& -\left(s_{1+}^{a+1}-1\right) \left(s_{1+}^{a+1}+1\right)\\& s_+^{2 b+1} (s_{1+}-s_+) (s_{1+} s_+-1)\big),
    \end{aligned}
\end{equation}

\begin{equation}
    \begin{aligned}
        M_2 =&s_+ s_{2+}^{2 b+2} \big(s_+^{2 a+1} \left(s_{1+}^{2 a+2} (s_{1+}-s_+)^2-(s_{1+} s_+-1)^2\right)\\&-\left(s_{1+}^{a+1}-1\right) \left(s_{1+}^{a+1}+1\right) \\&s_+^{2 b} (s_{1+}-s_+) (s_{1+} s_+-1)\big),
    \end{aligned}
\end{equation}

\begin{equation}
    \begin{aligned}
 M_3 =&s_{2+}^{2 c+1} \big(2 s_+^{2 a+1} \left(s_{1+}^{2 a+2} (s_{1+}-s_+)^2-(s_{1+} s_+-1)^2\right)\\&-\left(s_+^2+1\right) \left(s_{1+}^{a+1}-1\right) \left(s_{1+}^{a+1}+1\right)\\& s_+^{2 b} (s_{1+}-s_+) (s_{1+} s_+-1)\big),
    \end{aligned}
\end{equation}

\begin{equation}
    \begin{aligned}
M_4&=s_{2+}^{2 c+2} \big(\left(s_{1+}^{a+1}-1\right) \left(s_{1+}^{a+1}+1\right) s_+^{2 b+1} \\&(s_{1+}-s_+) (s_{1+} s_+-1) \\ &+s_+^{2 a} \left((s_{1+} s_+-1)^2-s_{1+}^{2 a+2} (s_{1+}-s_+)^2\right)\big), \\
    \end{aligned}
\end{equation}

\begin{equation}
    \begin{aligned}
 M_5 =& s_+ s_{2+}^{2 c} \big(\left(s_{1+}^{a+1}-1\right) \left(s_{1+}^{a+1}+1\right) s_+^{2 b}\\& (s_{1+}-s_+) (s_{1+} s_+-1)\\&+s_+^{2 a+1} \left((s_{1+} s_+-1)^2-s_{1+}^{2 a+2} (s_{1+}-s_+)^2\right)\big),
    \end{aligned}
\end{equation}

\begin{equation}
    \begin{aligned}
 M_6 =&s_{2+}^{2 b+1} \big(\left(s_+^2+1\right) \left(s_{1+}^{a+1}-1\right) \left(s_{1+}^{a+1}+1\right) \\&s_+^{2 b} (s_{1+}-s_+) (s_{1+} s_+-1)\\&+2 s_+^{2 a+1} \left((s_{1+} s_+-1)^2-s_{1+}^{2 a+2} (s_+-s_+)^2\right)\big).
    \end{aligned}
\end{equation}

\JV{The transmission probability is given by the absolute square $|\mathcal{T}|^2$, because the two leads have the same effective wave number.}

\subsection{Mott--Semiconductor Structure with Single Gate}
\label{app:w1w2}
\JV{For the heterostructure combining the two Mott insulating regions $U$ with the gate $V_m$  (see Fig.~\ref{fig:onegate}) the boundary conditions change. In the case of a half-filled background we find from Eq.~\ref{difference} the eight equations}
\begin{equation}
\label{eq:boundary_one}
\begin{aligned}
    \tilde{B} r_+ + \tilde{A} r_+^{-1}&=R s_+ + s_+^{-1}, \\
    1+R&=\tilde{A}+\tilde{B},\\
    \tilde{A} r_+^{a+1}+\tilde{B} r_+^{-(a+1)}&=C m_+^{a+1}+D m_+^{-(a+1)},\\
    C m_+^a+Dm_+^{-a}&=\tilde{A}r_+^a+\tilde{B} r_+^{-a},\\
    \tilde{E} r_+^b+\tilde{F} r_+^{-b}&= C m_+^b + Dm_+^{-b},\\
    \tilde{E} r_+^{c+1}+\tilde{F} r_+^{-(c+1)}&=\mathcal{T} s_+^{c+1},\\
    \mathcal{T} s_+^c &= \tilde{E} r_+^c +\tilde{F} r_+^{-c}, \\
    C m_+^{b+1}+D m_+^{-(b+1)}&=\tilde{E} r_+^{b+1}+\tilde{F} r_+^{-(b+1)}.
\end{aligned}
\end{equation}
\JV{where we introduced the abbreviations abbreviations $\{\tilde{A},\tilde{B},\tilde{E},\tilde{F}\}=\frac{E-V-U/2}{E-U-V}\{A,B,E,F\}$. Inserting the \textit{ansatz} \ref{eq:mottstructureansatz} yields the transmission amplitude}
\begin{equation}
\begin{aligned}
    \mathcal{T}(E,k^\|)=& \frac{\left(m^2-1\right) \left(r^2-1\right)^2 \left(s^2-1\right) \left(-m^{a+b}\right) r^{a+b+c+1}}{s^{c+1} \big( W_1 + W_2 \big)}.
       \end{aligned}
\end{equation}
In this, the two long expressions read
\vspace{1cm}
\begin{widetext}
\begin{equation}
    \begin{aligned}
    W_1& = -\left(m^{2 a} \left(r^{2 a+3} (r-s)+r s-1\right) \left(r^{2 b} (r s-1)+r^{2 c+1} (r-s)\right)\right)\\
    &-r m^{2 a+2} \left(r \left(r^{2 a} (r-s)+s\right)-1\right) \left(r^{2 b+1} (r s-1)+r^{2 c} (r-s)\right)\\
    &+r m^{2 b} \left(r \left(r^{2 a}
   (r-s)+s\right)-1\right) \left(r^{2 b+1} (r s-1)+r^{2 c} (r-s)\right)\\
   &+m^{2 b+2} \left(r^{2 a+3} (r-s)+r s-1\right) \left(r^{2 b} (r s-1)+r^{2 c+1} (r-s)\right)    \\
    W_2&=m^{2 a+1} \left(r^{2 b+1} (r s-1) \left(\left(r^2+1\right) r^{2 a+1} (r-s)
      +2 r s-2\right)+r^{2 c} (r-s) \left(2 r^{2 a+3} (r-s)+\left(r^2+1\right) (r s-1)\right)\right)\\
      &+m^{2 b+1} \left(r^{2 b+1} (r s-1) \left(-\left(\left(r^2+1\right) r^{2 a+1}
   (r-s)\right)-2 r s+2\right)-r^{2 c} (r-s) \left(2 r^{2 a+3} (r-s)+\left(r^2+1\right) (r s-1)\right)\right)\end{aligned}
\end{equation}
\end{widetext}

%\vfill
\bibliography{untitled3.bib}

\end{document}